\documentclass[aps,prd,showpacs,twocolumn,
superscriptaddress]{revtex4-1}

\usepackage{graphicx}
\usepackage{dcolumn}

\usepackage{amsmath}
\usepackage{amssymb}
\usepackage{bm}
\usepackage{color}
\usepackage[normalem]{ulem}
\usepackage[dvipsnames]{xcolor}
\usepackage{hyperref}
\hypersetup{
	colorlinks=true, 
	linkcolor=blue,  
	citecolor=cyan,  
}

\begin{document}
	
\title{Charged particle motion around a quasi-Kerr compact object immersed in an external magnetic field }

\author{Bakhtiyor Narzilloev}
\email{nbakhtiyor18@fudan.edu.cn}
	\affiliation{Center for Field Theory and Particle Physics and Department of Physics, Fudan University, 200438 Shanghai, China }

	\author{Ahmadjon Abdujabbarov}
	\email{ahmadjon@astrin.uz}
	\affiliation{Center for Field Theory and Particle Physics and Department of Physics, Fudan University, 200438 Shanghai, China }
	\affiliation{Ulugh Beg Astronomical Institute, Astronomicheskaya 33,
		Tashkent 100052, Uzbekistan }

	\author{Cosimo Bambi}
	\email{bambi@fudan.edu.cn}
	\affiliation{Center for Field Theory and Particle Physics and Department of Physics, Fudan University, 200438 Shanghai, China }

	\author{Bobomurat Ahmedov}
	\email{ahmedov@astrin.uz}
	\affiliation{Ulugh Beg Astronomical Institute, Astronomicheskaya 33,
		Tashkent 100052, Uzbekistan }

	\date{\today}

\begin{abstract}
We explore the electromagnetic fields around a quasi-Kerr compact object assuming it is immersed in an external asymptotically uniform magnetic field. Using the Wald method, components of the electromagnetic field in orthonormal basis have been obtained. We explore the charged particle motion around
deformed Kerr compact objects in the presence of external
asymptotically uniform magnetic fields. Using the Hamilton-Jacobi
equation, we obtain the effective potential expression  for
the charged particle surrounding a quasi-Kerr compact object immersed in
an external magnetic field. It is also derived the dependence of
innermost stable circular orbits (ISCOs) from the magnetic and deformation parameters for charged particles motion around a
rotating quasi-Kerr compact object. Comparison with ISCO radius measurements has provided the constraint to the deformation parameter as $\epsilon \gtrsim -0.012$.  The center of mass (CM) energy of the colliding particles in several physically interesting cases has been studied.

\end{abstract}

\maketitle

\section{Introduction}

Astrophysical black holes are believed to be a Kerr one, which has two parameters: its mass $M$ and rotation parameter $a$. However, there are many attempts to construct extension of Kerr black holes introducing extra parameters and parametrizations,~see, for example,~\cite{Newman63,Zimmerman89,Glampedakis06b,
Johannsen10,Johannsen11,Johannsen13,Konoplya16,
Cardoso14,Sen92,Konoplya16,Rezzolla2014}. 
The presence of nonvanishing electric charge in the spacetime metric has been tested by different scenarios in~\cite{Grunau11,Zakharov94,Stuchlik02, Pugliese10, Pugliese11b, Pugliese11, Patil12}. The properties of the black holes with brane charge have been studied in~\cite{Turimov17, Whisker05, Majumdar05, Liang17, Li15}. Other extension of the Kerr solution is the solution with gravitomagnetic charge~\cite{Liu11,Zimmerman89, Morozova09, Aliev08, Ahmedov12, Abdujabbarov11, Abdujabbarov08}. Various works are dedicated to test the axial symmetric metrics with the deformation parameters~\cite{Bambi17c,Rayimbaev15, Rayimbaev16, Bambi11b, Chen12, Bambi12,Bambi13e,Bambi17b,Cao18}. 

The authors of the Ref.~\cite{Glampedakis06b} proposed the deviations from the Kerr metric considering an approximate solution of Einstein vacuum equations and introducing the leading order deviation coming from the spacetime quadrupole moment. Different physical properties of these quasi-Kerr black holes have been studied in~\cite{Psaltis12, Liu12b}. Our recent work has been devoted to studying the weak lensing near the compact object with nonzero quadrupole momentum~\cite{Chakrabarty18}.

The magnetic field is very important in many astrophysical scenarios related to compact objects. Rotating neutron stars, having their own magnetic field, can be observed as pulsars~\cite{Landau32,Ginzburg1964,Rezzolla01c,Rezzolla01d}. However, the black holes, according to no-hair theorem,  may not create their own magnetic field~\cite{Ginzburg1964,Anderson70,Price72,Thorne72a}. The accretion disc around the rotating black holes can provide the magnetic field in the vicinity of the latter. The properties of the electromagnetic field structure around rotating black holes immersed in the external magnetic field were first initiated by Wald~\cite{Wald74}. The dipolar magnetic field configuration around a black hole created by circular electric current has been studied by Petterson~\cite{Petterson74}. After that the properties of the electromagnetic field structure around rotating black holes have been considered by various authors. The role of the magnetic field through magnetic Penrose process has been studied by~\cite{wagh85,Dhurandhar84b,Dhurandhar84,Dhurandhar83,Chellathurai86,Bhat85,Dadhich18}. The similar scenarios in alternative/modified theories of gravity have been studied in~\cite{Abdujabbarov08,Abdujabbarov10,Abdujabbarov11a,Abdujabbarov13a,Abdujabbarov13b,Abdujabbarov14,Stuchlik14a}. The charged particle motion around black holes immersed in external magnetic field has been studied by various authors~\cite{Aliev02,Frolov10,Frolov12,Karas12a,Stuchlik16,Kovar10,Kovar14,Kolos17,Shaymatov18} . 

In this work our main purpose is to study the electromagnetic field structure around rotating quasi-Kerr black holes with nonvanishing quadrupole momentum. The paper is
organaized as follows:  Sect.~\ref{sect2} is devoted to study the electromagnetic field components around quasi-Kerr compact objects immersed in external asymptotically uniform magnetic field. The charged particle motion around the quasi-Kerr compact object is studied in~Sect.~\ref{sect3} in the presence of magnetic field. In Sect.~\ref{sect4} we analyze the particle acceleration process around a compact object with nonzero quadrupole moment and in the presence of magnetic field. In Sect.~\ref{sect5}, we summarize the obtained results. 

\section{Compact object immersed in magnetic field \label{sect2}}

The metric for quasi-Kerr compact object is given by (for $G=c=1$, and with metric signature $(-+++)$) \cite{Glampedakis06b}

\begin{equation}\label{linelement}
ds^2=g_{00}dt^2+g_{11}dr^2+g_{22}d\theta^2+g_{33}d\phi^2+2g_{03}dtd\phi\, ,
\end{equation}
where
\begin{eqnarray}
\nonumber g_{00}&=&-\left(1-\frac{2Mr}{\Sigma}\right)+\epsilon
(1-3 \cos^2\theta)
\\\nonumber
&&\times \left(\frac{F_1
(1-\frac{2Mr}{\Sigma})^2}{1-\frac{2M}{r}}+\frac{4 a^2 F_2 M^2
\sin^2\theta} {\Sigma^2}\right) \, ,
\\\nonumber
g_{11}&=&\frac{\Sigma}{\Delta}+\epsilon \frac{F_1
(1-\frac{2M}{r})\Sigma^2 (1-3 \cos^2\theta)}{\Delta^2}\, ,
\\\nonumber
g_{22}&=&\Sigma-\epsilon \frac{F_2 \Sigma^2 (1-3 \cos^2\theta)}
{r^2}\, ,
\\\nonumber
g_{33}&=& \left(r^2+a^2+\frac{2a^2Mr
\sin^2\theta}{\Sigma}\right)\sin^2\theta
\\\nonumber
&&+ \epsilon\,(1-3 \cos^2\theta) \sin^2\theta
\left[\frac{4a^2F_1M^2r^2 }{(1-\frac{2M}{r}) \Sigma^2} \right.
\\\nonumber
&&\left. -\frac{F_2}{r^2} \left(a^2+r^2+\frac{2a^2Mr
\sin^2\theta}{\Sigma}\right)^2 \right]\, ,
\\\nonumber
g_{03}&=&-\frac{2aMr}{\Sigma} \sin^2\theta+\epsilon (1-3
\cos^2\theta) \sin^2\theta
\\\nonumber
&&\times\left[\frac{2aF_1Mr (1-\frac{2Mr}{\Sigma})
}{(1-\frac{2M}{r}) \Sigma} \right.
\\
&&\left. +\frac{2aF_2M }{r\Sigma}(a^2+r^2+\frac{2a^2Mr
\sin^2\theta}{\Sigma})\right]\, ,
\label{metric}
\end{eqnarray}
with $\Sigma =r^2+ a^2 \cos ^2\theta$, 
$\Delta =r^2-2 M r+a^2$ and

\begin{eqnarray}
F_1&=&\frac{5 (M-r) \left(2 M^2+6 M r-3 r^2\right)}{8 M r (r-2 M)}
\\\nonumber
&&-\frac{15 r (r-2 M) }{16 M^2}\ln\left(\frac{r}{r-2 M}\right)\, ,
\\
F_2&=&\frac{5 \left(2 M^2-3 M r-3 r^2\right)}{8 M r}
\nonumber\\
&&+\frac{15 \left(r^2-2 M^2\right)}{16 M^2}\ln \left(\frac{r}{r-2 M}\right)\, ,\label{metfunct}
\end{eqnarray}
the constant $\epsilon$ in the expression 
(\ref{metric}) indicates the small contribution to 
the quadrupole moment $Q$ of the compact object with the total mass $M$ as
\begin{eqnarray}
Q&=&-M(a^2+\epsilon M^2)\ ,
\end{eqnarray}
where the deformation parameter $\epsilon$ 
might take either positive $\epsilon > 0$ or 
negative $\epsilon < 0$
values~\cite{Glampedakis06b,Chakrabarty18}. One can easily see that the case when
$\epsilon = 0$ the spacetime metric (\ref{linelement})--(\ref{metfunct}) describes the spacetime of a 
Kerr black hole.

In order to find the vector potential 
of the electromagnetic field in the vicinity of the compact object
we will follow the Wald method that assumes the black hole is immersed 
in a uniform magnetic field~\cite{Wald74,Aliev02,Abdujabbarov08,Abdujabbarov10,Stuchlik14a}. Here we use the existence 
in this spacetime of a timelike Killing vector, 
${\xi}^{\alpha}_{(t)}=\partial x^{\alpha}/\partial
t$, and a spacelike one, ${\xi}^{\alpha}_{(\phi)}=\partial
x^{\alpha}/\partial \phi$, which are responsible for the
stationarity and axial symmetry of the spacetime geometry~ (\ref{linelement})--(\ref{metfunct})
which satisfies the Killing equations
\begin{eqnarray}
{\xi}_{\alpha ; \beta}+{\xi}_{\beta ; \alpha}=0 \ ,
\end{eqnarray}
and according to the Wald method ~\cite{Wald74} the 
solution of the vacuum Maxwell's equations $\Box A^{\mu}=0$ for
the vector potential $A_{\mu}$ of the electromagnetic field in the
Lorentz gauge can be written as 
\begin{equation}
A^{\alpha}=C_{1}\xi^{\alpha}_{(t)}+C_{2}\xi^{\alpha}_{(\varphi)}\ .
\end{equation}
The constant $C_{2} = B/2$, where the compact object is immersed
in the uniform magnetic field $\mathbf{B}$ that is aligned 
along its rotating axis. The remaining constant $C_1$ can be
found from the asymptotic properties of spacetime~(\ref{linelement})--(\ref{metfunct}) at infinity.
Indeed in order to find the remaining constant one can use the
electrical neutrality of the compact object
$ Q^*=0$
evaluating the value of the integral through the spherical surface
at the asymptotic infinity. Then one can easily get the value of
constant $C_1=aB$.

The contravariant components of the vector potential $A^{\mu}$ 
of the electromagnetic field will take the following form
\begin{eqnarray}
 A^{0}=a B \, ,\,  A^{1}=A^{2}=0\ \, ,\,  A^{3}=\frac{1}{2}\,B \, .
\label{potential}
\end{eqnarray}
Now one can easily find the covariant components of the 
vector potential using $ A_{\mu}=g_{\mu\nu} A^{\nu} $ 
for metric~(\ref{linelement})--(\ref{metfunct}).

\begin{widetext}
\begin{eqnarray}
\label{covA}
A_0&=&a B \left\{\epsilon  \left(1-3 \cos ^2\theta \right) \left[\frac{F_1 r}{r-2 M}\left(\frac{2 M r}{\Sigma }-1\right)^2-\frac{4 a^2 F_2 M^2 \sin
   ^2\theta }{\Sigma ^2}\right]+\frac{2 M r}{\Sigma }-1\right\}\nonumber
   \\\nonumber
   &&+\frac{a B M r}{\Sigma } \sin ^2\theta  \left\{\epsilon  \left(1-3 \cos ^2\theta \right)
   \left[\frac{F_2 \mathcal{R}}{r^2 \Sigma }-\frac{F_1 r}{r-2 M} \left(\frac{2 M r}{\Sigma }-1\right)\right]-1\right\}\ ,
\\
A_1&=&A_2=0\ ,
\\
A_3&=&\frac{2 a^2 B M r \sin ^2 \theta }{\Sigma } \left(1-3 \cos ^2 \theta \right) \left\{\epsilon  \left[\frac{F_2 \mathcal{R}}{r^2 \Sigma }-\frac{F_1 r}{r-2 M}\left(\frac{2 M
	r}{\Sigma }-1\right)\right]-1\right\}\nonumber
   \\\nonumber
   &&+\frac{1}{2} B \sin ^2 \theta  \left\{\epsilon  \left(1-3 \cos ^2 \theta \right)
   \left[\frac{4 a^2 M^2 r^3 F_1 \sin ^2 \theta }{\left(r-2 M\right) \Sigma ^2}-\frac{F_2 \mathcal{R}^2}{\Sigma^2 r^2}\right]+\frac{\mathcal{R}}{\Sigma }\right\}\ ,
\end{eqnarray}
\end{widetext}
where $ \mathcal{R}= 2 a^2 M r \sin^2 \theta+(a^2+r^2) \Sigma $. 

The components of the electromagnetic fields can be found using following expressions in curved spacetime.
\begin{eqnarray}
\label{fields}
E_\alpha &=& F_{\alpha \beta} u^\beta\ ,
\\
B^\alpha &=& \frac{1}{2} \eta^{\alpha \beta \sigma \mu} F_{\beta \sigma} u_\mu \ ,
\end{eqnarray}
with 
\begin{equation}
F_{\alpha \beta}=A_{\beta; \alpha}-A_{\alpha; \beta}\ ,
\end{equation}
where $F_{\alpha \beta}$, $u^\beta$ and $\eta^{\alpha \beta \sigma \mu}$ are the electromagnetic field tensor, the four velocity of the observer and the Levi-Civita tensor, respectively.

In the ZAMO (zero angular momentum observer) system the four velocity is defined as follows:

\begin{eqnarray}
\label{uzamo}
(u_\alpha)_{ZAMO}&=&\left(-\frac{1}{\cal U },0,0,0\right)
\\
(u^\alpha)_{ZAMO}&=&\left({\cal U },0,  0, -\frac{2 a M r}{\Delta  \Sigma  {\cal U}}\right)
\end{eqnarray}
with
\begin{eqnarray}
{\cal U} &=&\sqrt{\frac{F_1 r \epsilon  \left(1-3 \cos ^2\theta \right)}{r-2 M}+\frac{\mathcal{R}}{\Delta  \Sigma }}
\end{eqnarray}

Finally, the orthonormal components of the electromagnetic field components read as


%
%
\begin{widetext}
\begin{eqnarray}
E^{\hat{r}}&=&-\frac{a B M \mathcal{P}_1}{2 \Sigma ^2 \sqrt{\mathcal{R}}}+\frac{a B \sqrt{\mathcal{R}}}{4 \Delta  \Sigma^2} \epsilon  \left\{2 \Delta  \Sigma  \left[\frac{N_2+\frac{2 N_3 \left(1-3 \cos ^2\theta \right)}{(r-2
   M)^2}}{\Sigma ^3}-\frac{M r \sin ^2\theta  \left(N_4+\frac{2 N_5}{r^3}\right)}{\mathcal{R}}\right]+N_1\right\} \label{e1}
   \\
E^{\hat{\theta}}&=&\frac{a B M r  \mathcal{P}_2}{\Sigma ^2 \sqrt{\Delta  \mathcal{R}}}\sin 2 \theta +\frac{a B}{4 r \Sigma }
   \epsilon  \sqrt{\frac{\mathcal{R}}{\Delta }} \left\{K_1+\frac{K_2 \sin 2 \theta }{\Sigma ^2 (2 M-r)}-\frac{2 M r^2}{\mathcal{R}} \left[K_3+\frac{2
   (K_4 \sin 2\theta +K_5)}{r^2}\right]\right\} \label{e2}
   \\
B^{\hat{r}}&=&\frac{B \mathcal{P}_3   }{2 \sqrt{\frac{\mathcal{Q} \Sigma }{\Delta }}}\sin
   2 \theta+\epsilon  \left\{\frac{1}{2} B \sqrt{\frac{\Sigma }{\Delta }} \left[\frac{\Delta  \left(\frac{R_4 \sin 2 \theta
   }{r^2}+R_5\right)}{\sqrt{\mathcal{Q}}}+\frac{  \mathcal{P}_3 \sin 2 \theta
   \left(\frac{R_3}{\sqrt{\mathcal{Q}} 2 r^2 \Sigma \left(2 M  - r \right)}+R_2\right)}{\Sigma }\right]+R_1\right\} \label{b1}
   \\
B^{\hat{\theta}}&=&\frac{B \mathcal{P}_4 \Delta \sin^2 \theta
    }{\Sigma 
   \sqrt{\Sigma \mathcal{Q}}}+\epsilon  \left\{\frac{1}{2} B \sqrt{\Sigma } \sin ^2\theta  \left[\frac{(2 D_2+D_3) \left(8 a^2 M r^2 - \mathcal{P}_4\right)}{\Sigma
   ^2}+\frac{D_4-D_5}{\sqrt{\mathcal{Q}} r^3}\right]-D_1\right\} \label{b2}
\end{eqnarray}
\end{widetext}
where $\mathcal{Q} = \left(\mathcal{R} - 2 a^2 M r - 2 M r^3 \right) \mathcal{R} \sin^2\theta\ ,$ and explicit forms of functions $ \mathcal{P}_i $, $ i=1..4 $ with $ N_j, K_j, R_j $ and $ D_j $, $ j=1..5 $ are given in Appendix \ref{appendix}.

\begin{figure*}[t!]
\begin{center}
\includegraphics[width=0.31\linewidth]{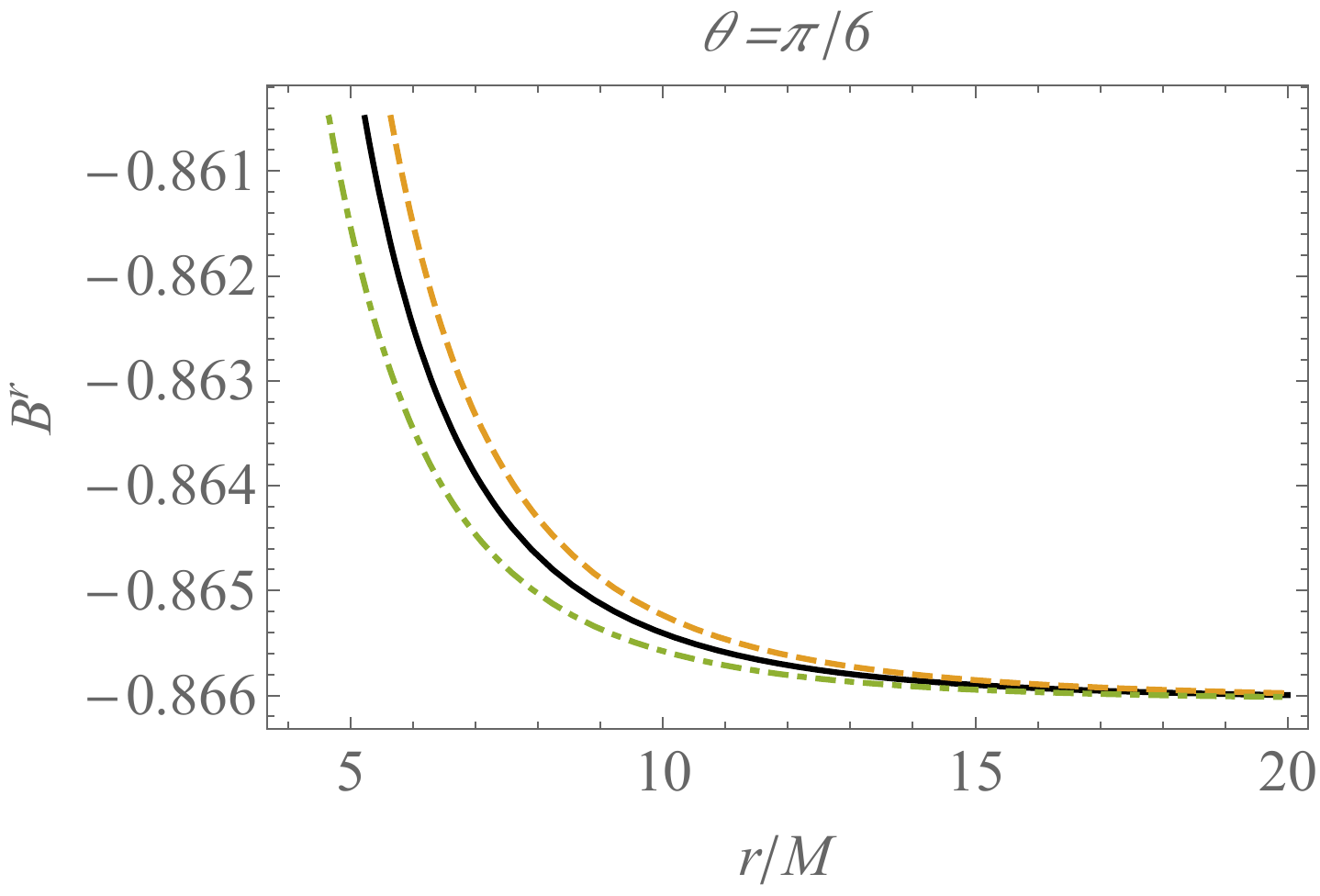}
\includegraphics[width=0.31\linewidth]{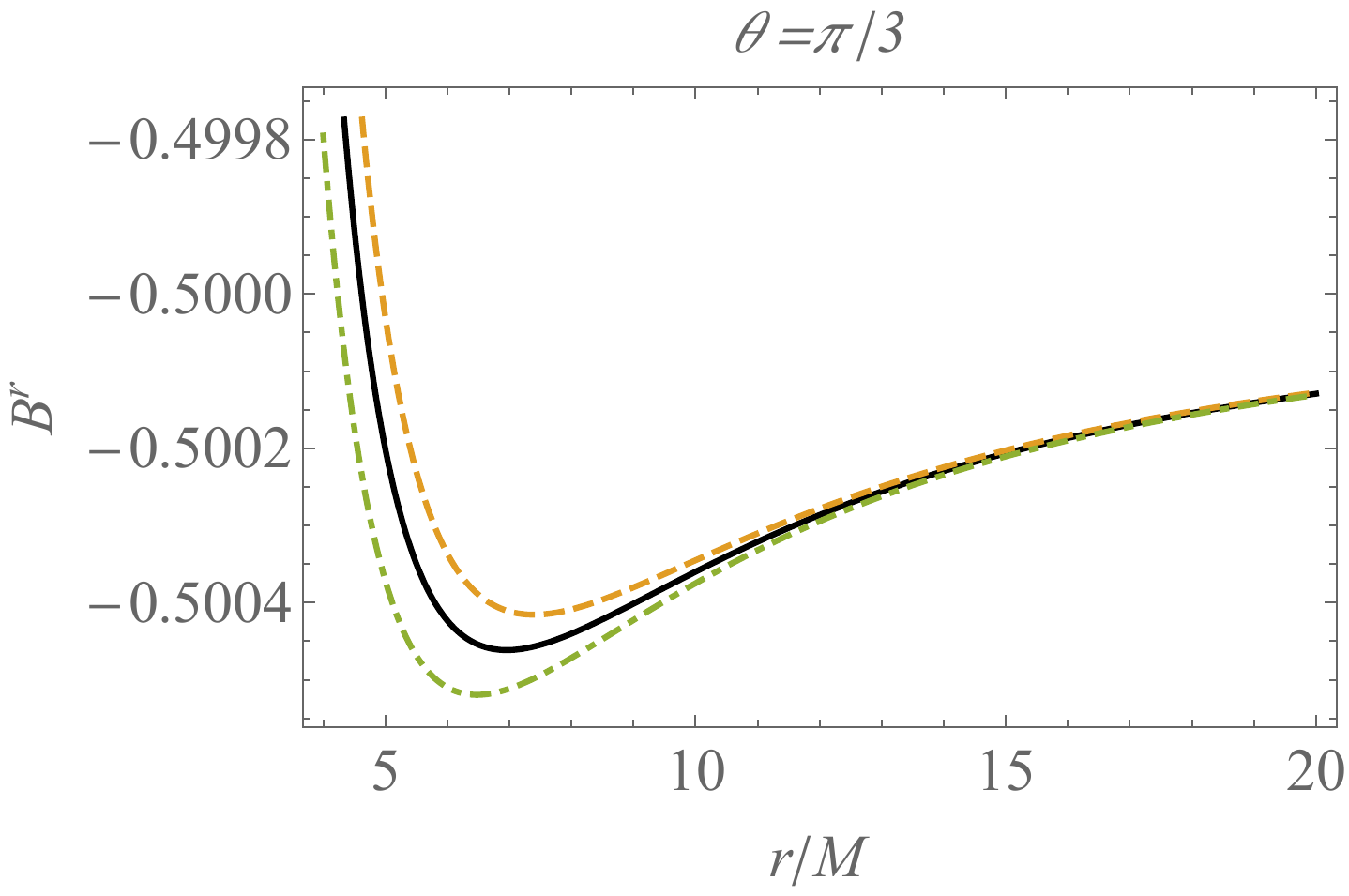}
\includegraphics[width=0.31\linewidth]{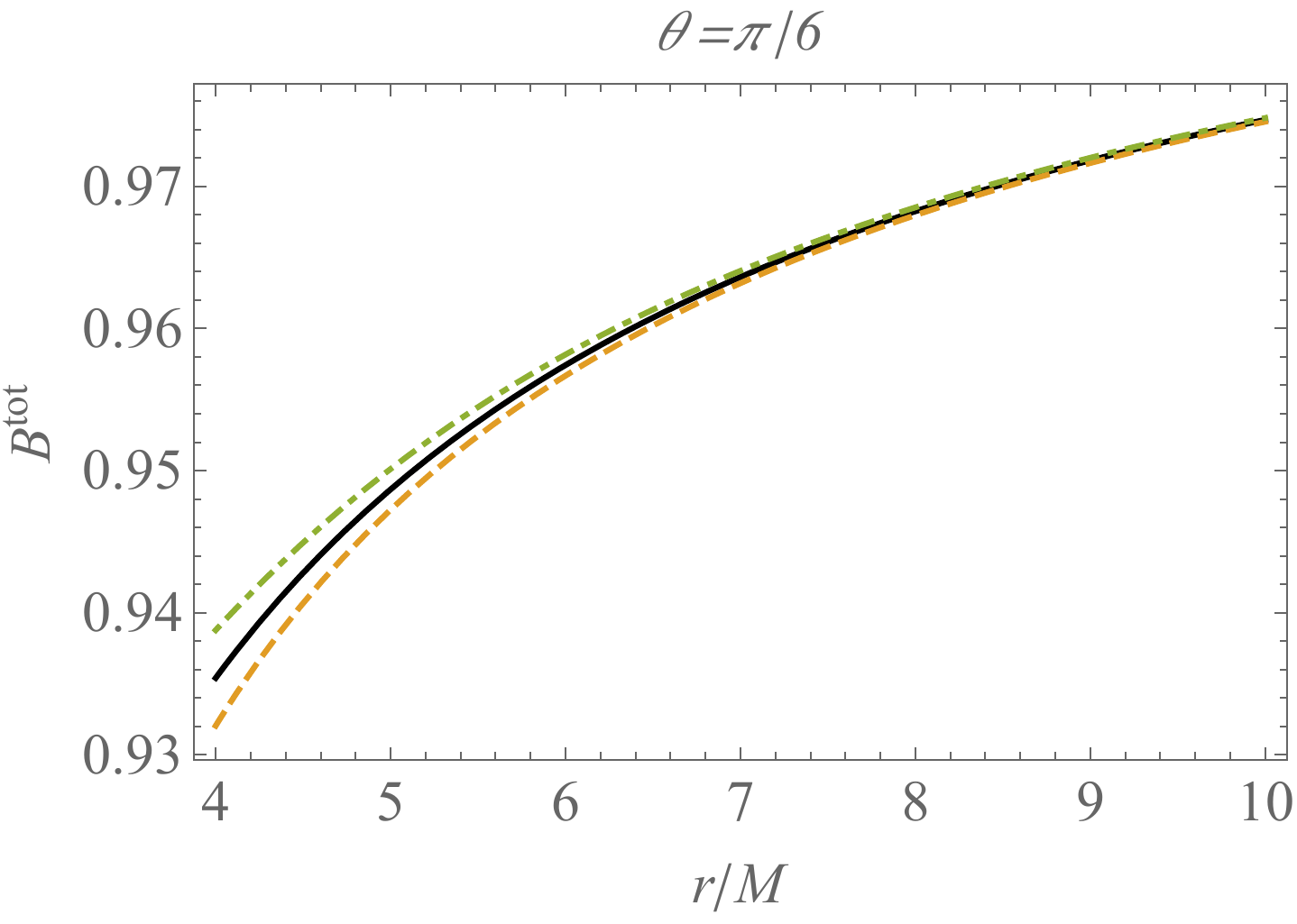}

\includegraphics[width=0.31\linewidth]{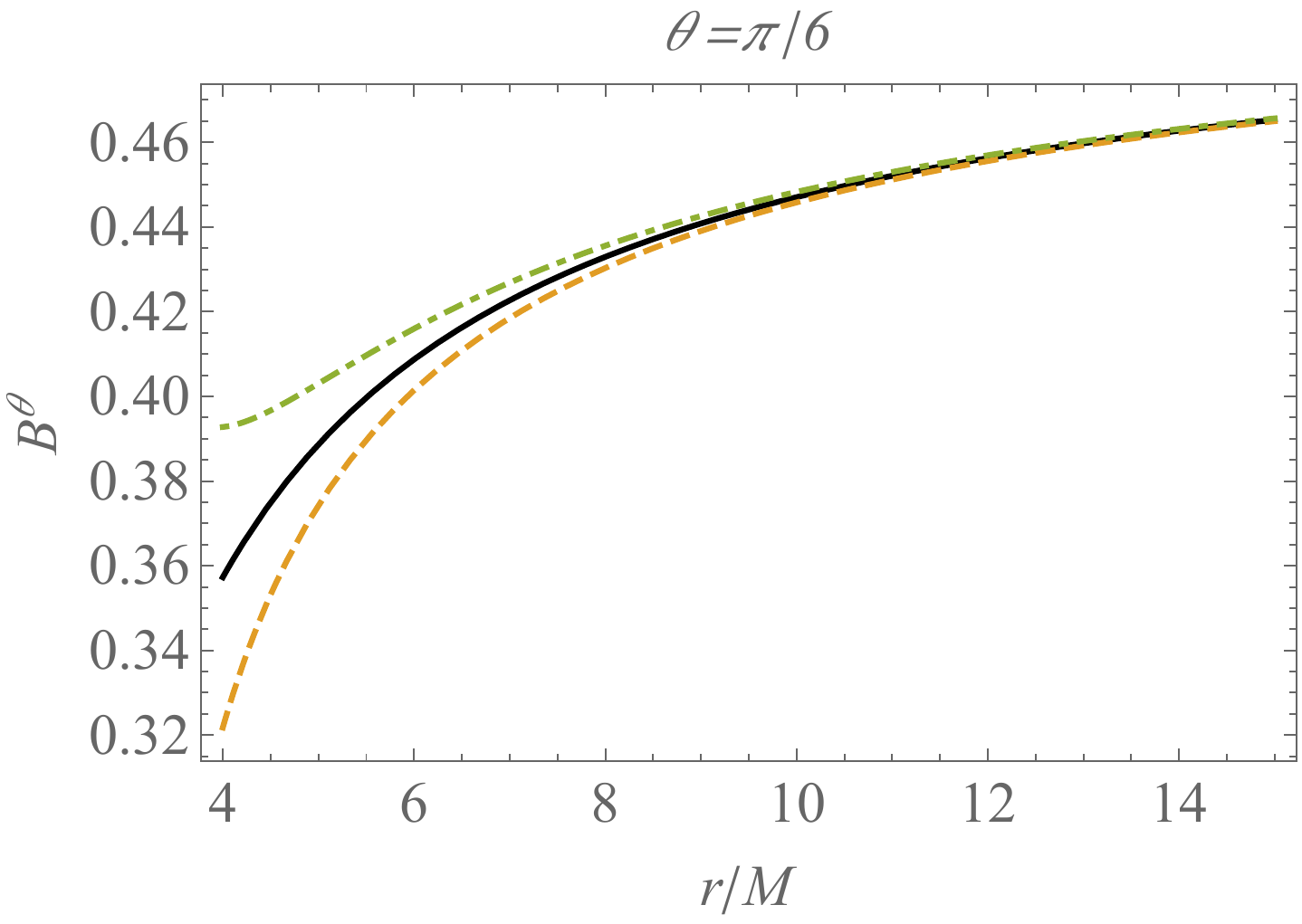}
\includegraphics[width=0.31\linewidth]{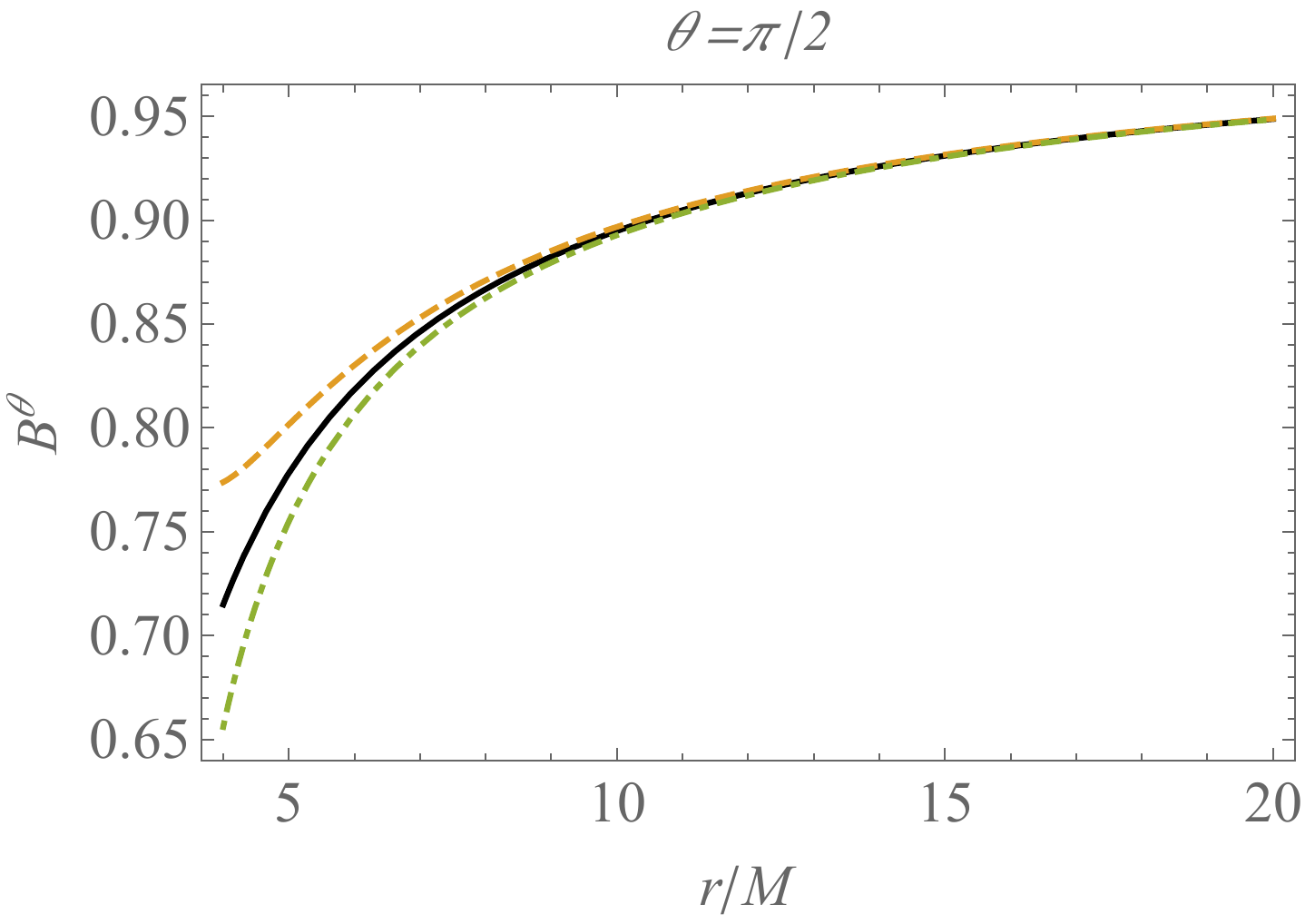}
\includegraphics[width=0.31\linewidth]{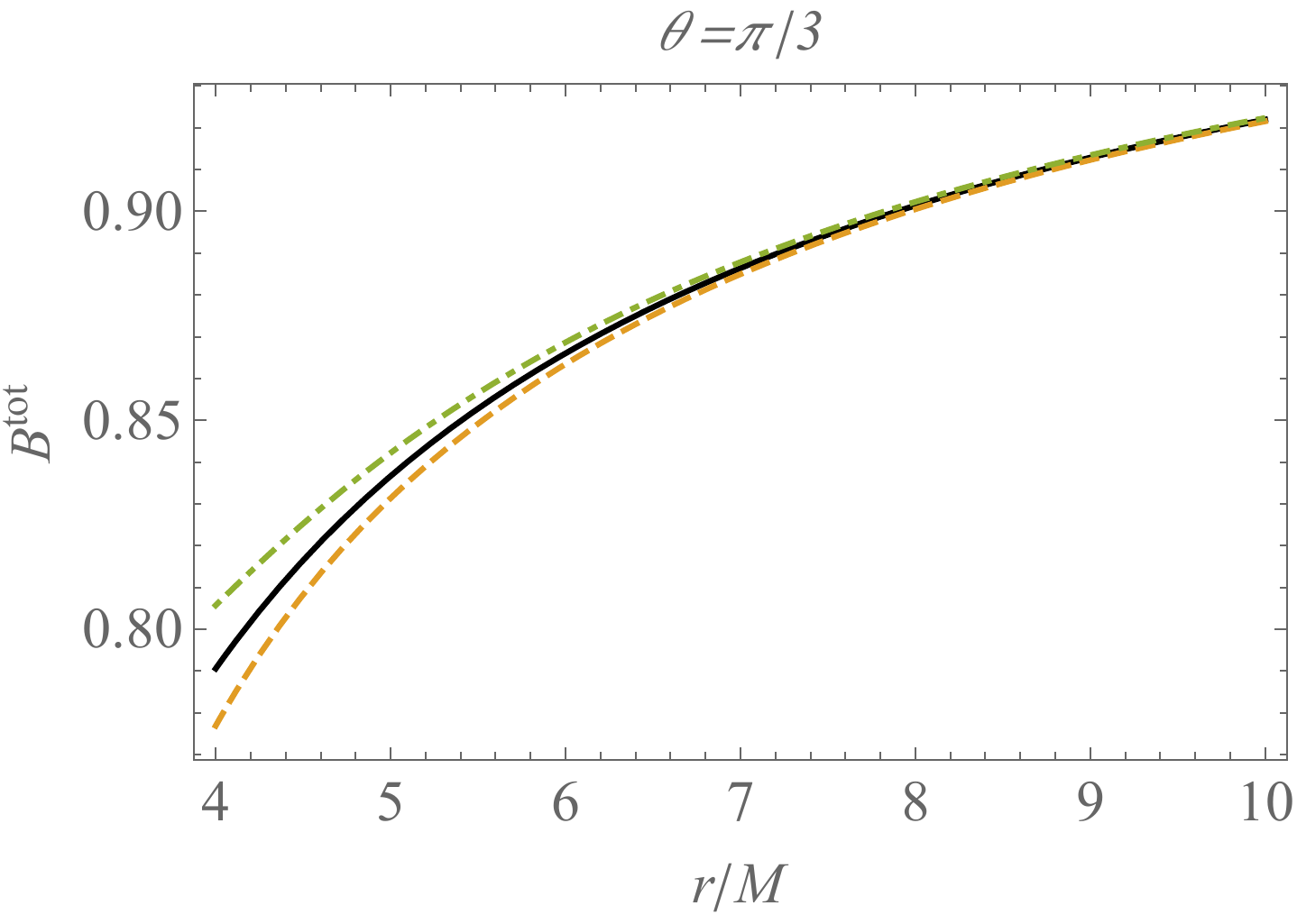}

\includegraphics[width=0.31\linewidth]{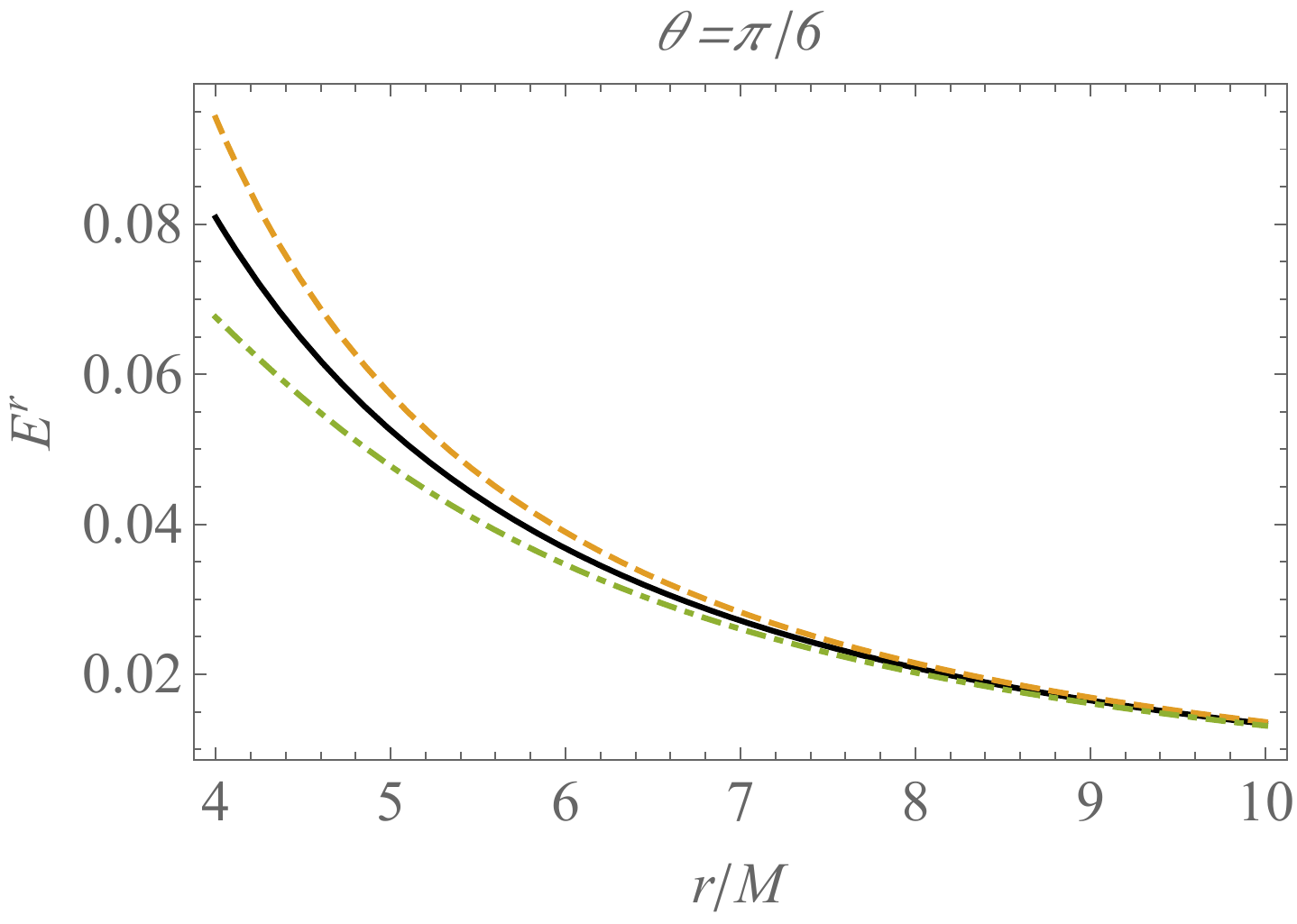}
\includegraphics[width=0.31\linewidth]{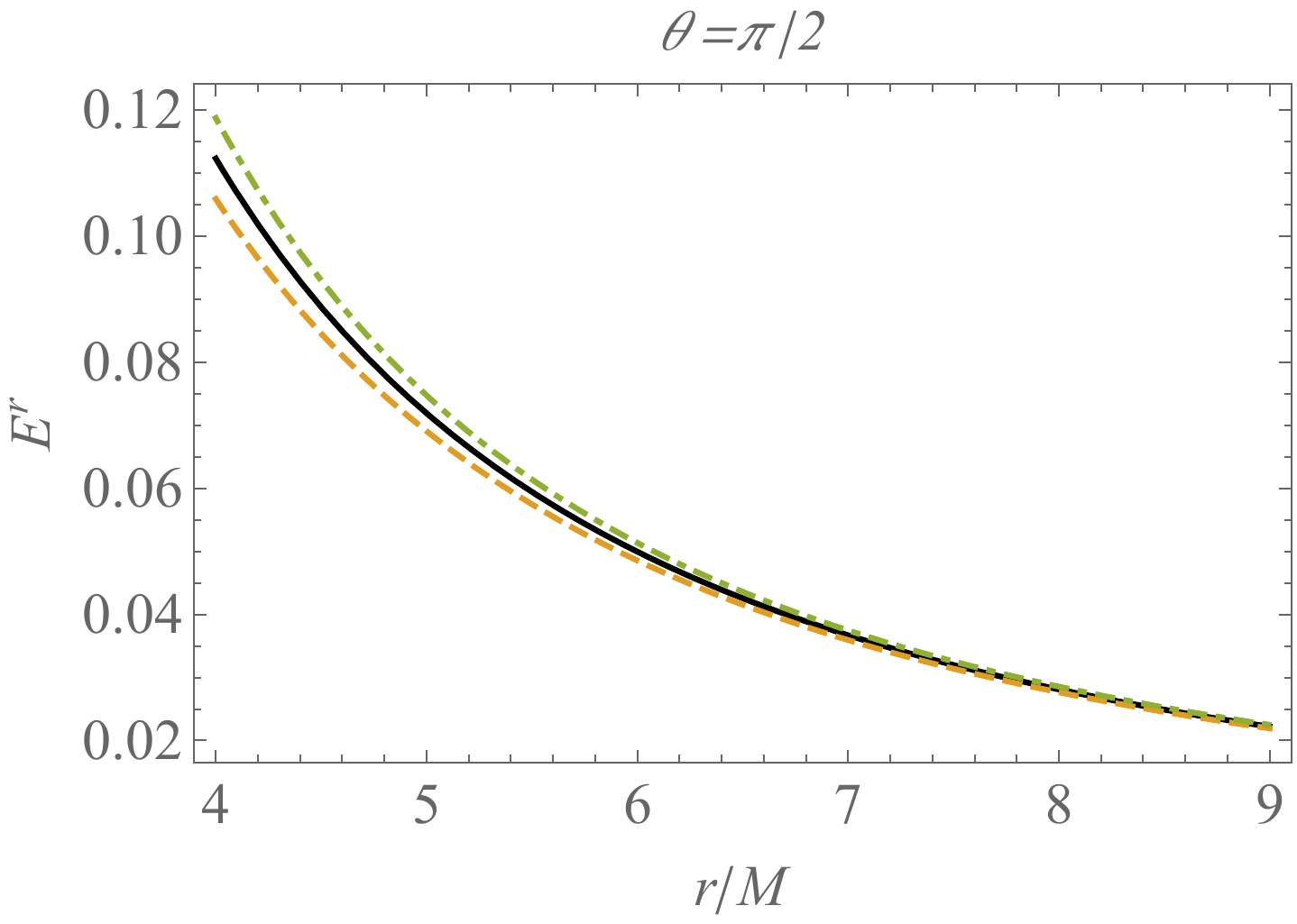}
\includegraphics[width=0.31\linewidth]{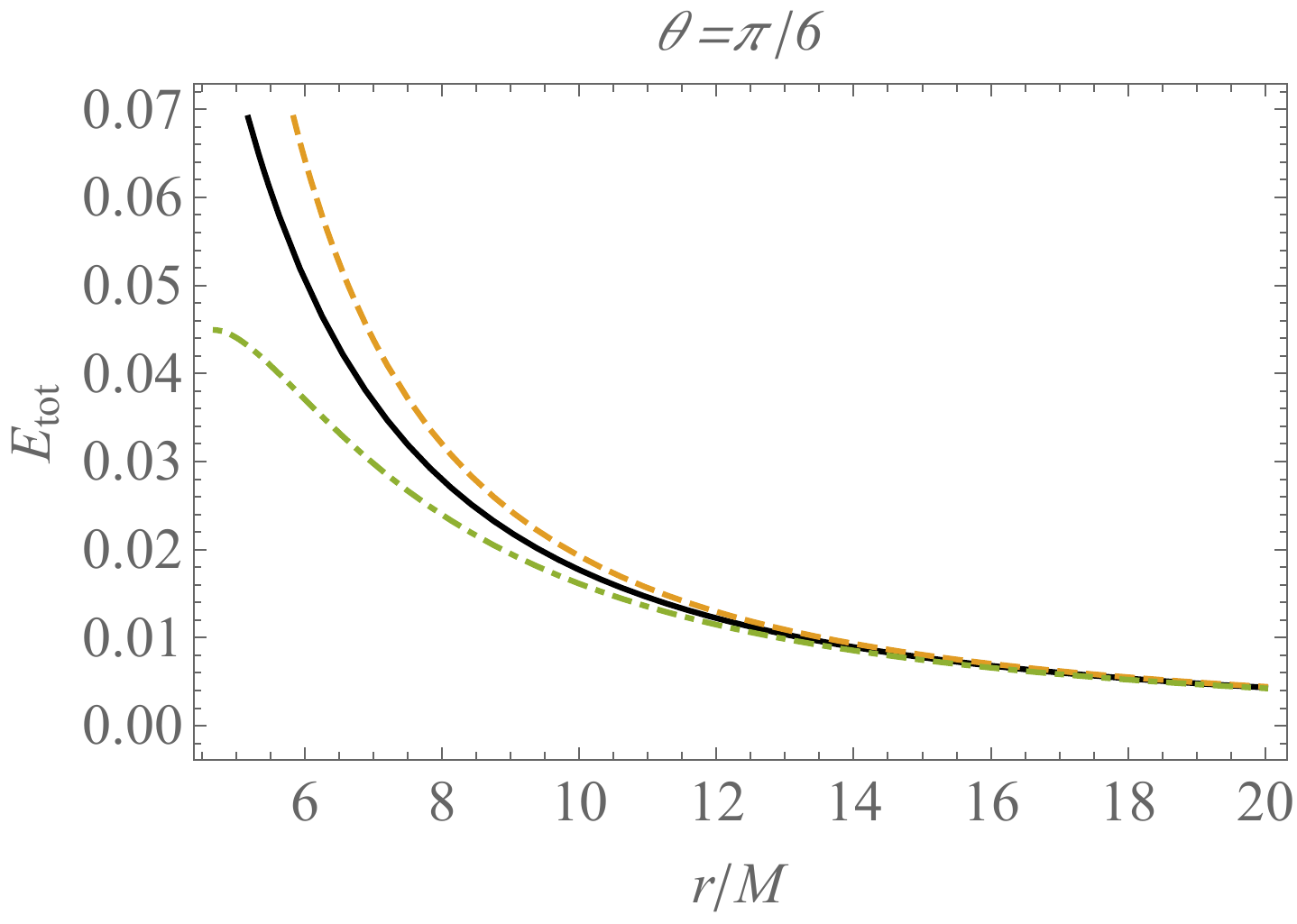}

\includegraphics[width=0.31\linewidth]{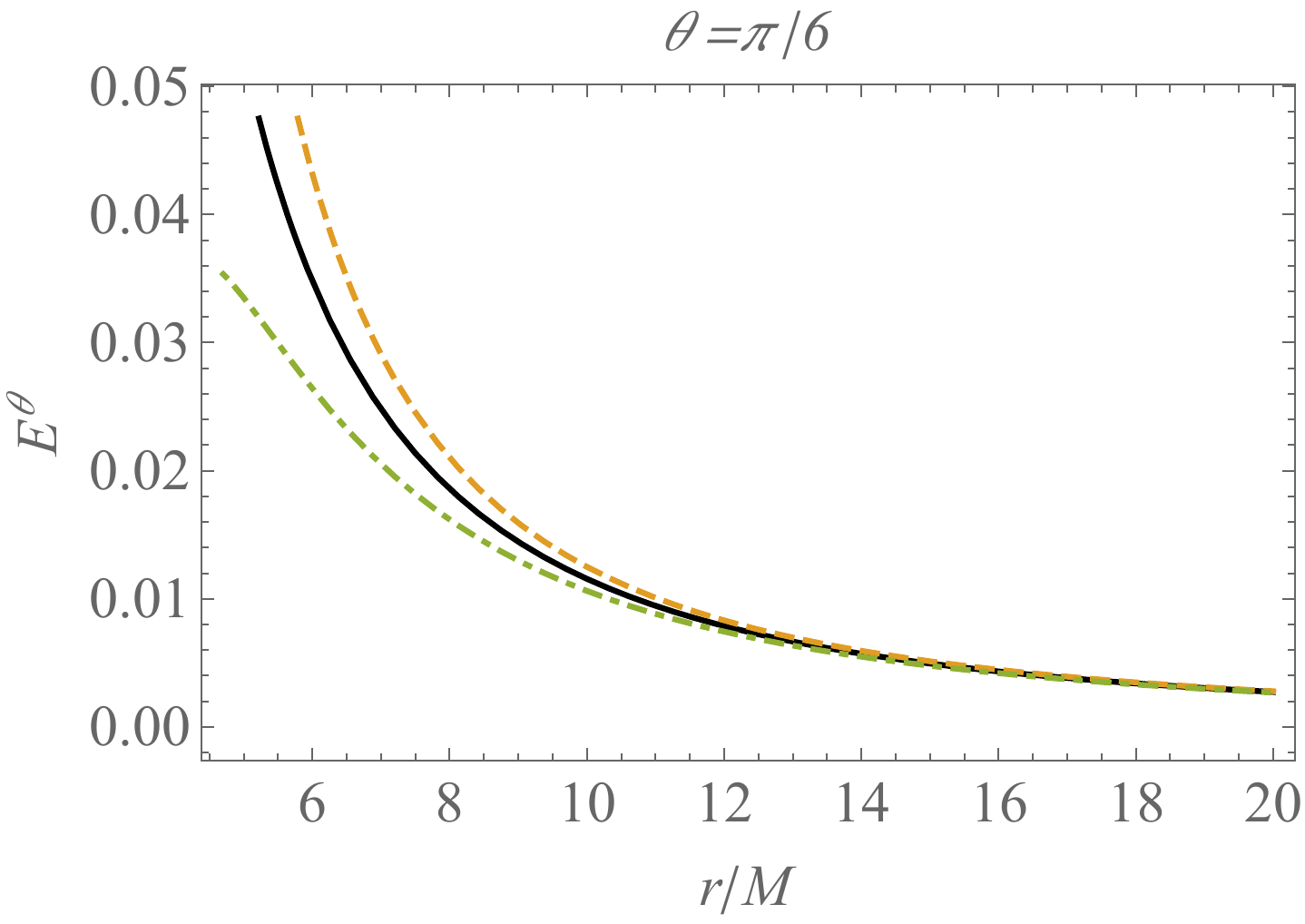}
\includegraphics[width=0.31\linewidth]{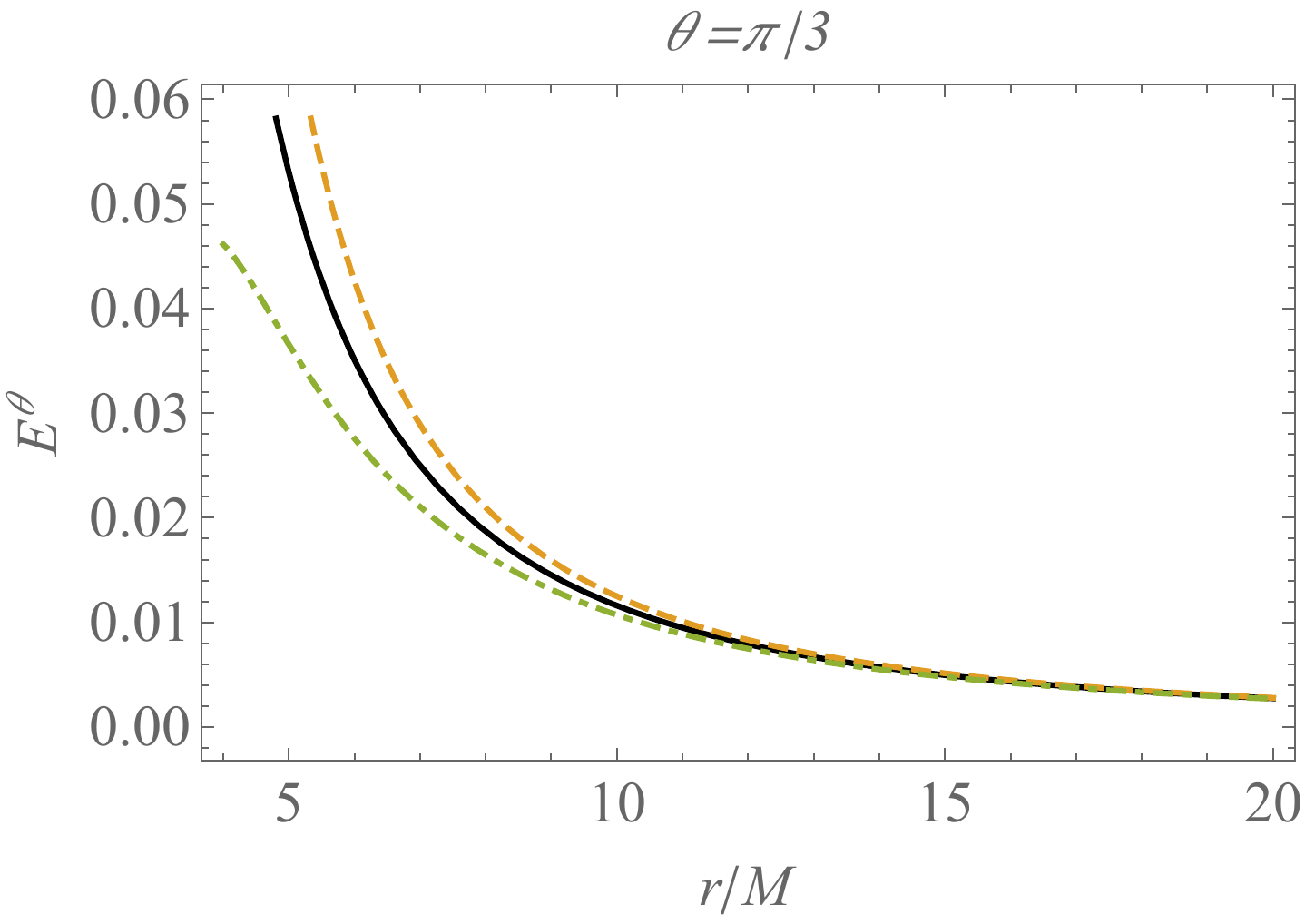}
\includegraphics[width=0.31\linewidth]{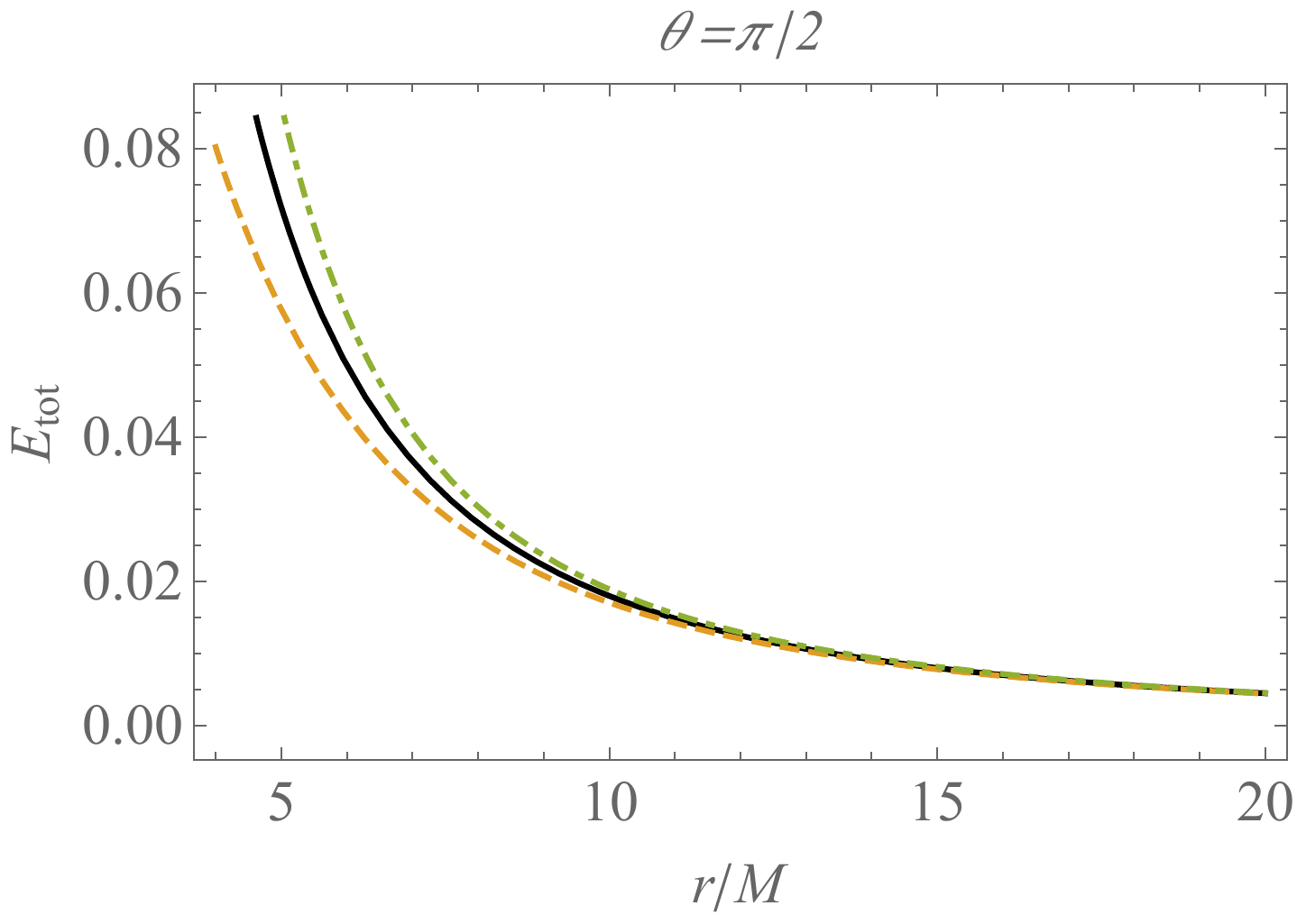}
\end{center}

\caption{The radial dependence of the electric and magnetic fields for different values of $\theta$. In all graphics we set $B=1$ and $M=1$ for simplicity. Black solid lines correspond to $\epsilon=0$ while dashed orange ones and green dot-dashed ones to $\epsilon=0.5$ and $\epsilon=-0.5$, respectively.
\label{fig_1}}
\end{figure*}

In the case of the rotating quasi - Kerr compact object one can obtain simplified expressions for the fields in the linear or quadratic approximation  ${\cal O}(a^2/r^2, \epsilon)$.
\begin{eqnarray}\label{20}
 E^{\hat{r}}&=& \frac{a B M}{r^2} \left[  \cos ^2\theta -3+\frac{3 M^2 \epsilon (1-3 \cos ^2 \theta)}{r^2} \right] \ ,
\\\nonumber
\\\label{21}
E^{\hat{\theta}} &=& \frac{a B M}{r^2} \sin 2 \theta \left[2 - \frac{2 M}{r} - \frac{ 3 M^2 \epsilon }{r^2}  \right.
\nonumber\\
&& \left. + \frac{3 (a^2  \cos ^2 \theta - M^2) }{r^2}\right] \ ,
\\
 B^{\hat{r}} &= &B \cos \theta \left[1+\frac{a^2}{2 r^2} \sin ^2 \theta \right.
\nonumber\\
&& \left.-\frac{a^2 M}{2 r^3} \left(5+3 \cos 2\theta \right) -\frac{3 M^3 \epsilon }{r^3} \sin ^2 \theta  \right] \ ,
\\\nonumber
\\
B^{\hat{\theta}} &=& B \sin \theta \Big[1-\frac{M}{r}-\frac{M^2 + a^2 \cos ^2\theta }{2
   r^2}+ \frac{M}{4 r^3} \Big.
   \nonumber\\
&& \Big. \times    \left\{a^2 (9+5 \cos 2\theta)- 3 M^2 \epsilon (1+3 \cos 2\theta)-2 M^2)\right\}\Big]. \nonumber\\
\end{eqnarray}

It can be seen from the equations (\ref{20}) and (\ref{21})  that in the
linear approximation in $\epsilon$ and $a$, electric field does not have any contribution from $\epsilon$. In the limit of flat spacetime, i.e., for $M/r \rightarrow 0$, expressions~(\ref{e1}) - (\ref{b2}) give
\begin{eqnarray} \label{newton limit}
E^{\hat{r}} = E^{\hat{\theta}}=0 ,\quad
B^{\hat{r}} = B \cos \theta , \quad B^{\hat{\theta}} = B \sin \theta \ .
\end{eqnarray}
As expected, expressions (\ref{newton limit}) coincide with the
solutions for the homogeneous magnetic field in the
Newtonian spacetime. This can be see from the plots of expressions (\ref{e1}) - (\ref{b2}) in Fig.\ref{fig_1}. Indeed, we can see that for the large distances the absolute values of the components of the electric field tend to zero while the components of the magnetic field tend to the corresponding values of $\sin\theta$ and $\cos\theta$ for chosen angles $\theta$ as in (\ref{newton limit}), when B=1.

\section{Charged particle motion around magnetized compact object \label{sect3}}

In this section we will consider the equation of motion of charged
particles in the background spacetime of rotating compact object with the metric given in~(\ref{linelement})--(\ref{metfunct}).
We are aimed at investigating the particle 
motion in the spacetime of a quasi-Kerr compact object immersed
in a uniform magnetic field. In order to describe the charged particle motion we use the Hamilton-Jacobi equation which can be expressed as
\begin{eqnarray}
g^{\alpha\beta} \left(\frac{\partial S}{\partial x^\alpha} - q A_\alpha\right)
\left(\frac{\partial S}{\partial x^\beta} - q A_\beta\right) = -m^2 \ ,
\label{HJ}
\end{eqnarray}
where $m$ and $q$ are the mass and charge of the test particle, respectively.

Due to the existence of two Killing vectors $\xi_{(t)}^\alpha$ and $\xi_{(\phi)}^\alpha$ the action of charged particle around the compact object can be described
as follows
\begin{eqnarray}
S=-Et+L\phi +S_1(r,\theta) \label{action}\, ,
\end{eqnarray}
where $E$ and $L$ are the energy and the angular momentum of the
charged particle, respectively. { It is worthwhile to note that
 from the symmetry of the problem it is clear that the circular
orbits are possible in the equatorial plane and  the
magnetic field is also oriented toward the spin of black
hole being perpendicular to equatorial plane at each point. Since we do not assume the force free condition   
 the magnetic field being perpendicular to the  equatorial plane will force the charged particle to move on this plane (see for example \cite{Aliev89,Aliev02,Frolov12b,Frolov12}). 
Substituting the action (\ref{action})  into the equation of motion of charged particle (\ref{HJ})
one can get the equation for unseparable part of the action. It is not possible to separate variables  in the general case. 
However, for the motion in equatorial plane ($\theta = \pi/2$) the equation
of motion maybe separable in spacetimes having symmetries and one may proceed the 
calculations for the equatorial motion (see~\cite{Aliev02}).} Inserting (\ref{action}) to
(\ref{HJ}) and after making  calculations in equatorial plane
$(\theta=\pi/2)$ one can easily find the equation for the radial part
of motion which corresponds to the
radial component of covariant 4-momentum of the charged particle $(p_r = \partial S_r/\partial r)$. 
Radial contravariant component of the momentum can be obtained 
multiplying the metric (\ref{linelement}) with covariant momentum. On the 
other hand, 
\begin{equation}\label{urdvij}
\left(\frac{dr}{d\tau}\right)^2=f(r)={\cal E}^2-1-2V_{\rm eff}^2\ ,
\end{equation}
where $\tau$ is the proper time of the test particle and 
$V_{\rm eff}^2$ is the square of the effective potential that can be written as 
\begin{equation}
V_{\rm eff}^2=V_{(K)}^2 + V_{(q)}^2\ .
\end{equation}
The first term corresponds to the Kerr one and the second indicates a small deviation taking place in quasi-Kerr spacetime. Solving equations (\ref{HJ}) and (\ref{urdvij}) we can obtain the following expressions for each term 
\begin{widetext}
\begin{eqnarray}
V_{(K)}^2&=&\frac{1}{8 r^3}\bigg\{a^4 b^2 (2 M-3 r)+8 a^3 b {\cal E} r-2 a^2 \bigg[r \left(b \left(b \left(2 M^2-4 M r+r^2\right)+2 {\cal L}\right)-2\right)+2 {\cal E}^2 (2 M+r)\bigg] 
\nonumber\\
&& +8
   a {\cal E} \bigg[b r^2 (r - 2 M)+2 {\cal L} M\bigg]-2 b^2 M r^4+b^2 r^5-4 b {\cal L} r^2 (r-2 M)+4 {\cal L}^2 (r-2 M)-8 M r^2 \bigg\} \ ,
\\
V_{(q)}^2&=&\frac{5 \epsilon}{128 M^2 r^7 \left(a^2+r (r-2 M)\right)}  \Bigg\{r^4 \left[2 M (M-r) \left(2 M^2+6 M r-3 r^2\right)-3 r^2 (r-2 M)^2 \ln \frac{r}{r-2 M}\right] \nonumber
\\\nonumber
&&  \times \bigg[a^4 b^2 (3 r-2
   M)-8 a^3 b {\cal E} r+2 a^2 \left(r \left(b \left(b \left(2 M^2-4 M r+r^2\right)+2 {\cal L}\right)-2\right)+2 {\cal E}^2 (2 M+r)\right)  
\\\nonumber   
&&  .   -8 a {\cal E}
   \left(b r^3 -2br^2 M+2 {\cal L} M\right)+r^2 \big[\left(b^2 r^2+4\right) (2 M-r)+4 {\cal E}^2 r\big]+4 b {\cal L} r^2 (r-2 M)+{\cal L}^2 (8 M-4
   r)\bigg] 
\\\nonumber   
&& \Big.   -\frac{1}{(r-2
   M)^2}\left[a^2+r^2-2 Mr\right]^2 \Bigg[(r-2 M)^2 \left[2 M \left(2 M^2-3 M r-3 r^2\right)+3 r \left(r^2-2 M^2\right) \ln
   \frac{r}{r-2 M}\right]
\\\nonumber   
&&  \times \bigg[a^2 b (r-2 M)+b r^3+2 {\cal L} r\bigg] \bigg[b (a^2 r-2 a^2 M+r^3)-2 {\cal L} r\bigg]+4 r^3 \bigg[3 r^2 (r-2
   M)^2 \ln \frac{r}{r-2 M} 
\\\nonumber   
&&    -2 M (M-r) \left(2 M^2+6 M r-3 r^2\right)\bigg] (a b r-abM+{\cal E} r) (a b r-abM-{\cal E} r)\Bigg]\Bigg\} \ ,
\end{eqnarray}
\end{widetext}
where we have introduced new notations ${\cal E}=E/m$ and ${\cal L}=L/m $ 
 and dimensionless magnetic parameter $b = qB/m$ which characterizes the 
cyclotron frequency of the charged particle.

\begin{figure*}[t!]
\begin{center}
a.
\includegraphics[width=0.45\linewidth]{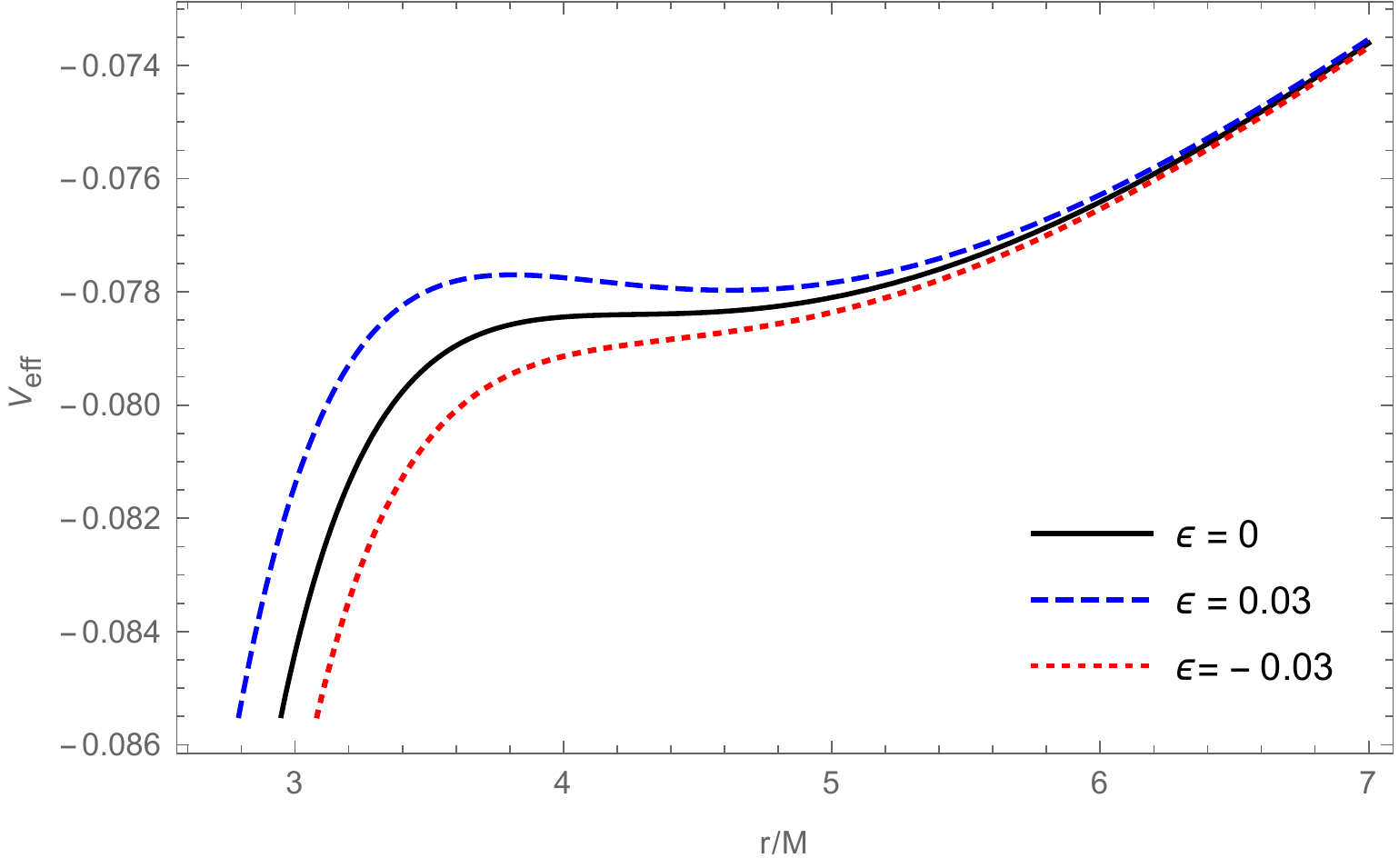}
b.
\includegraphics[width=0.45\linewidth]{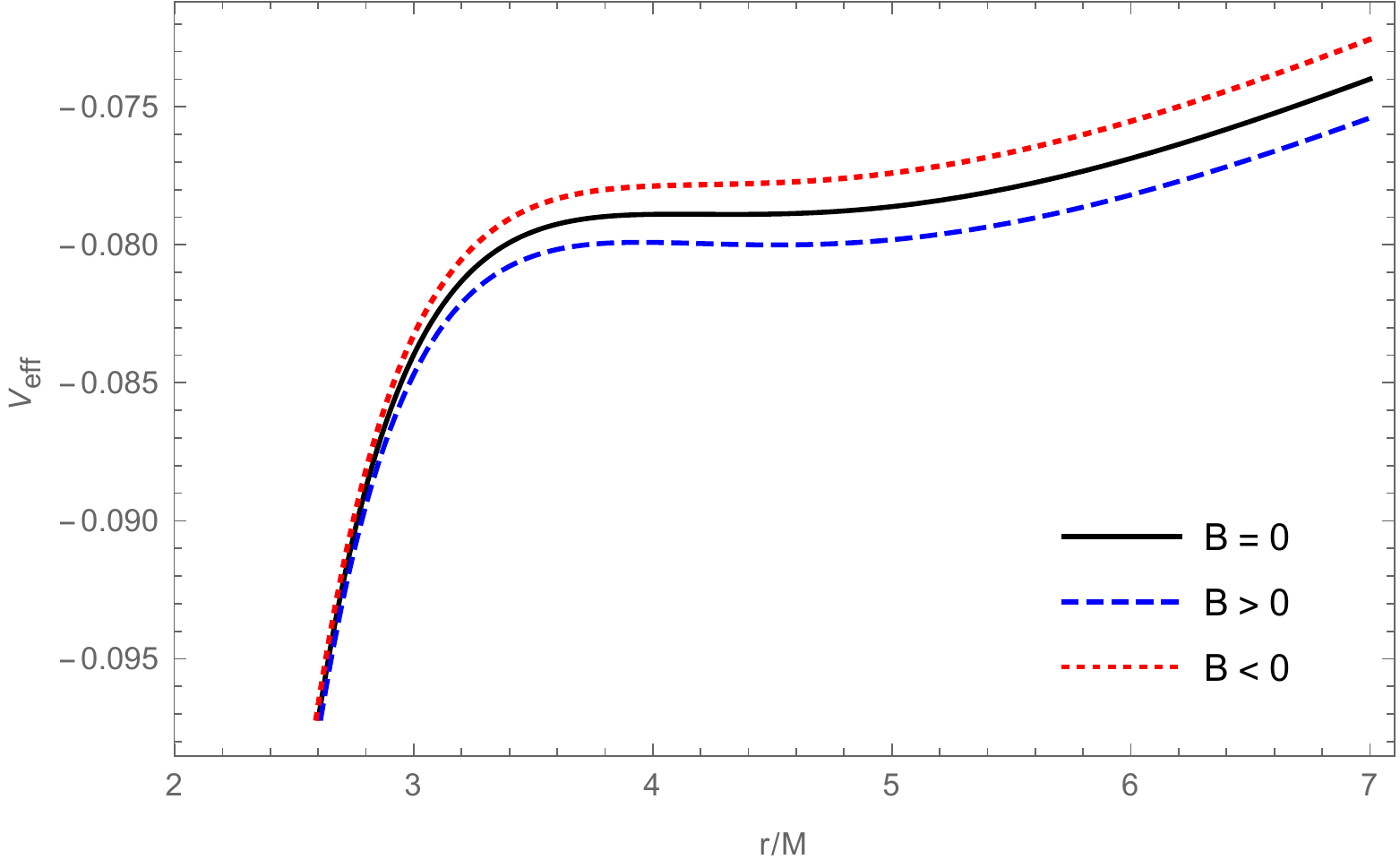}
\end{center}
\caption{The radial dependence of the effective potential of
radial motion of charged particle around rotating quasi-Kerr compact object in equatorial plane.  The figures correspond to the case of
slowly rotating compact object with $a=0.5$. a. without external magnetic field and b. in the presence of external magnetic field when $\epsilon=0.005$.\label{fig2}}
\end{figure*}

In Fig.~\ref{fig2} 
the radial dependence of the effective 
potential for different 
values of the magnetic and deformation parameters is shown. 
It is worth to note that, points where the lines turn correspond to ISCO of the charged test particle. 

Now we consider circular orbits, especially the innermost stable circular orbit for the charged test particle in the spacetime of a quasi-Kerr compact object using the following conditions
\begin{eqnarray}
f(r)=0 \ , \label{min}
\\
f'(r)=0 \ , \label{circle}
\\
f''(r)=0 \ . \label{isco}
\end{eqnarray}
By solving equations (\ref{min}), (\ref{circle}), and (\ref{isco}) 
all together one can find numerical values of the magnitudes for the energy, 
the angular momentum and the radius of ISCO of the charged 
particle orbiting around a quasi-Kerr compact object immersed 
in an asymptotically uniform external magnetic field. Obtained results are shown in 
Table~\ref{1tab} and Table~\ref{2tab}. It is apparent from 
the tables that, increasing the value of the deformation parameter 
$\epsilon$ and the external magnetic field strength will reduce the ISCO radius. It is worth to note that as $\epsilon=0$, the results coincide with the results for the Kerr compact object as expected. 
%
%
One can obtain the approximated analytical expression for the ISCO radius of the charged particle moving around quasi-Kerr compact object immersed in an external uniform magnetic field  in the absence of rotation and in the linear approximation of deviation parameter $\epsilon$ with quadratic term of magnetic field parameter $b$ (for the case $M=1$)
%
\begin{eqnarray}
&&r_{_{\rm ISCO}}\approx6-  \left[1440 \log
   (3/2)-\frac{9325}{16}\right]\epsilon
\\ \nonumber  
   &&-18 \big[24+5   \left\{805-1971 \log (3/2)\right\}\epsilon\big] b^2\ .
\end{eqnarray}
{
It should be mentioned that in the 
TAB.~\ref{1tab}  the extremal case when $a\approx1$ is not included. 
It is because in the extreme case the ISCO calculated would be in the regions close to the event horizon of
black hole where the quasi-Kerr metric diverges at the positions 
 requiring to take into account terms of order ${\cal{O}}(\epsilon^2)$ (see ~\cite{Johannsen10}). Consequently in the 
 linear approximation of the metric (\ref{linelement}) we skip the extreme case limiting ourselves with the results obtained above. }

\begin{table}
\caption{\label{1tab} ISCO radius of the particles moving around the rotating quasi-Kerr compact object
(case of $b=0$). }
\begin{ruledtabular}
\begin{tabular}{ccccccccc}
$\epsilon$ & -0.006 & -0.003 & 0 & 0.003 & 0.006  \\
\hline
$a=0$ & 6.0063 & 6.0032 & 6.0 & 5.9968 & 5.9936
\\
\hline
$a=0.5$ & 4.2515 & 4.2423 & 4.233 & 4.224 & 4.2141
\\
\hline
$a=0.7$ & 3.4430 & 3.4189 & 3.3931 & 3.3655 & 3.3355
\\
\hline
$a=0.8$ & 3.0287 & 2.9754 & 2.9066 & 2.7964 & 2.7551
\\
\end{tabular}
\end{ruledtabular}
\end{table}
\begin{table}
\caption{\label{2tab} ISCO radius of the particles moving around the rotating quasi-Kerr compact object
(case of $a=0.5$). }
\begin{ruledtabular}
\begin{tabular}{ccccccccc}
 $\epsilon$ & -0.006 & -0.003 & 0 & 0.003 & 0.006 \\
\hline
$b=0$ & 4.2515 & 4.2423 & 4.233 & 4.2236 & 4.2141
\\
\hline
$b=0.05$ & 4.0567 & 4.0483 & 4.0399 & 4.0313 & 4.0226
\\
\hline
$b=0.2$ & 3.3061 & 3.29798 & 3.2897 & 3.2812 & 3.2725
\\
\hline
$b=0.5$ & 2.7418 & 2.7294 & 2.7162 & 2.7018 & 2.6860
\\
\end{tabular}
\end{ruledtabular}
\end{table}

\section{Particle collisions in the vicinity of a compact object \label{sect4}}

{ It is well known that supermassive black holes anchored in the most galaxies and observed as active galactic nuclei are one of the 
 most powerful source of energy emission in the Universe. There are different mechanisms 
 as Blandford-Znajek  and Penrose processes allowing to extract energy from a rotating black hole (see \cite{Hawking74,Hawking75,Hawking76,wagh85a,wagh85})). 
 In~\cite{Banados09} it has also been shown that center of mass energy of two colliding particles around a 
 rotating black hole increases exponentially in the near regions allowing to extract some part of the rotational energy of a black hole. 
 Consequently particles collision can be considered as one of the important phenomena in the
studying black hole energetics and we devote
this section to the investigation of particle collision in the black hole environment.} Here we study some simple cases of the collisional processes of test particles that could well represent the role of the deviation parameter and the external magnetic field added to the gravitational field of the quasi-Kerr compact objects.

\subsection{Collisions of neutral particles with opposite angular momentum \label{subsect1}}

In this subsection we calculate the central mass energy of two colliding neutral particles falling from infinity with zero initial velocity  and opposite angular momentum $L_1=-L_2=L$.  For simplicity we assume that the particles are moving on the equatorial plane ($\theta=\pi/2$) and have the same initial rest energy ($E_{1,2}=m$) at infinity. Under such assumptions the 4-velocities of the particles read 

\begin{eqnarray}
\nonumber \label{conU}
u_i^t&=&-g^{tt} + g^{t\phi} \frac{L_i}{m} \ ,\\
u_i^\phi&=&-g^{t\phi} + g^{\phi\phi} \frac{L_i}{m} \ ,\\\nonumber
u_i^\theta&=&0 \ ,
\end{eqnarray}
and $u_i^r$ can be found from the condition $g_{\alpha\beta} \  u_i^\alpha \ u_i^\beta=-1$ with $i=1,2$.

The CM energy of the system of test particles can be found from the relation $E^2_{CM}=-g_{\alpha\beta} \  p^{\alpha}_{\rm tot} \  p^{\beta}_{\rm tot}$. The square of the CM energy then reads

\begin{eqnarray} \label{CME}
\nonumber
E^2_{CM}&=&-m^2 g_{\alpha\beta} (u_1^\alpha+u_2^\alpha) (u_1^\beta+u_2^\beta)\\
&=& 2 m^2 (1-g_{\alpha\beta} u_1^\alpha u_2^\beta) \ .
\end{eqnarray}

or

\begin{widetext}
\begin{eqnarray}
&&\frac{E^2_{CM}}{2 m^2}=\frac{2 a^2 (M+r)+{\cal L}^2 (r-2 M)+2 r^2 (r-M)-{\cal Z}}{a^2 r+r^2 (r-2 M)}
\\\nonumber
 &+& \epsilon  \frac{-2 a^2 M \left[F_1 r^3+F_2 {\cal L}^2 (2 M-r)\right]+F_1 r^3 \left[{\cal L}^2 (r-2 M)-2 M r^2+{\cal Z}\right]+F_2
   {\cal L}^2 (r-2 M) \left[{\cal L}^2 (2 M-r)+2 M r^2+{\cal Z}\right]}{r^2 {\cal Z} (r-2 M)} \ ,
\end{eqnarray}
where $\mathcal{Z}=\sqrt{\bigg[2 M \left((a-{\cal L})^2+r^2\right)-{\cal L}^2 r\bigg]\bigg[2 M \left((a+{\cal L})^2+r^2\right)-{\cal L}^2 r\bigg]}$.
\end{widetext}

When one assumes such particles to collide at the turning point which is given by the following condition

\begin{eqnarray} \label{tpoint}
u^r(r_t,{\cal L})=0 \ ,
\end{eqnarray}
the specific angular momentum of colliding particles as a function of the turning point radius reads

\begin{widetext}
\begin{eqnarray}
{\cal L}(r_t)=\frac{\sqrt{2} {\cal G}-2 a M r_t^3}{r_t^3 (r_t-2 M)} + \frac{{\cal G} \epsilon  \left[-2 \sqrt{2} a^2 F_2 M r_t^3 (2 M+r_t)+8 a F_2 {\cal G} M+\sqrt{2} r_t^5 (2 M-r_t) (2 F_2 M-F_1 r_t)\right]}{4 M r_t^8 (r_t-2
   M)^2} \ ,
\end{eqnarray}
\end{widetext}
with ${\cal G}=\sqrt{M r_t^7 \left[a^2+r_t (r-2 M)\right]}$.

\begin{figure}
\begin{center}
\includegraphics[scale=0.5]{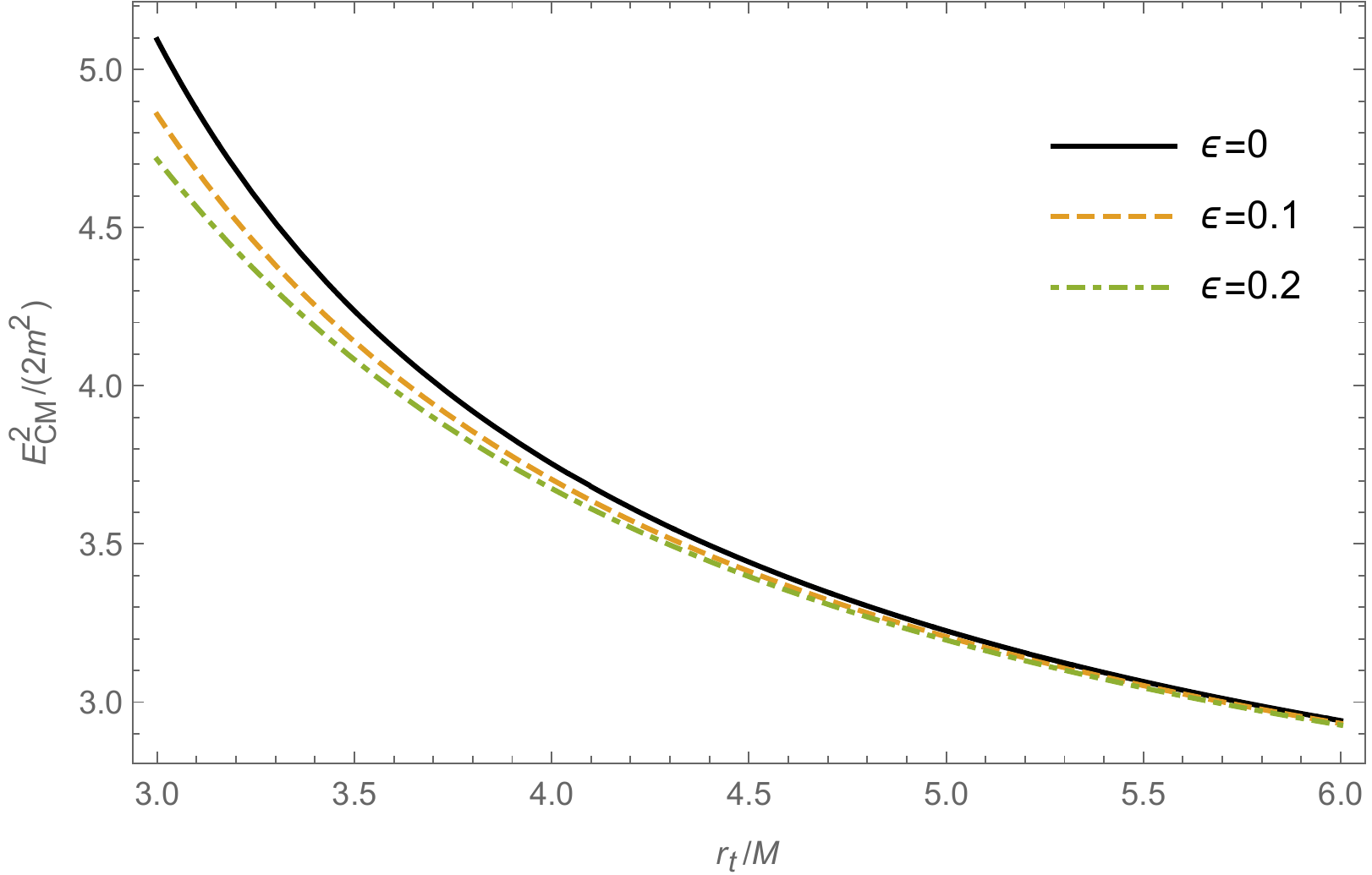}
\end{center}
\caption{The radial dependence of the CM energy of the colliding neutral particles for the different values of the deviation parameter $\epsilon$. The rotation parameter is taken $a=0.5$.}\label{fig3}
\end{figure}

 The behavior of the CM energy $E^2_{CM}/2m^2$  as a function
of the radius $r_t$ is illustrated in Fig. \ref{fig3}. It is apparent from the graph that in the absence of the deviation parameter, $\epsilon=0$, the CM energy of the particles goes up exponentially near to the compact object as in the case of Kerr one, whereas, in the presence of some deviation one can see significant differences in the shape of lines and can conclude that increasing of the deviation parameter reduces the CM energy of the particles.

\subsection{ Collision of charged particles on circular orbits with radially falling neutral ones }

In this subsection we focus on the collision of two particles where the first particle is charged and orbiting on a circular orbit around the compact object and the second one is electrically neutral and falling from infinity. To make our calculations easy to solve we set the following assumptions: 

(i) both particles have the same rest mass ($m_1=m_2=m$); 
	
(ii) rotation is absent ($a=0$); 
	
(iii) particles are moving on the same plane; 
	
(iv) the neutral particle is falling  radially from infinity.

The four velocity of the neutral particle that is falling from infinity reads
\begin{eqnarray}\nonumber
u_2^t&=&-g^{tt} \ ,\\
u_2^r&=&\sqrt{g^{rr} (-1-g^{tt})} \ ,\\\nonumber
u_2^\theta&=&u_2^\phi=0 \ .
\end{eqnarray}
The four velocity of the charged particle moving on circular orbit can be found using conditions (\ref{min}) and (\ref{circle})
\begin{eqnarray}\nonumber
u_1^t&=&-g^{tt} \mathcal{E}(r,b,\epsilon) \ ,\\
u_1^r&=&u_1^\theta=0 \ ,\\\nonumber
u_1^\phi&=&g^{\phi\phi} \mathcal{L}(r,b,\epsilon) \ ,
\end{eqnarray}
where $\mathcal{E}(r,b,\epsilon)$ and $\mathcal{L}(r,b,\epsilon)$ represent the energy and the angular momentum of the charged particle and they are derived solving equations(\ref{min}) and (\ref{circle}).

\begin{figure}
\begin{center}
\includegraphics[scale=0.55]{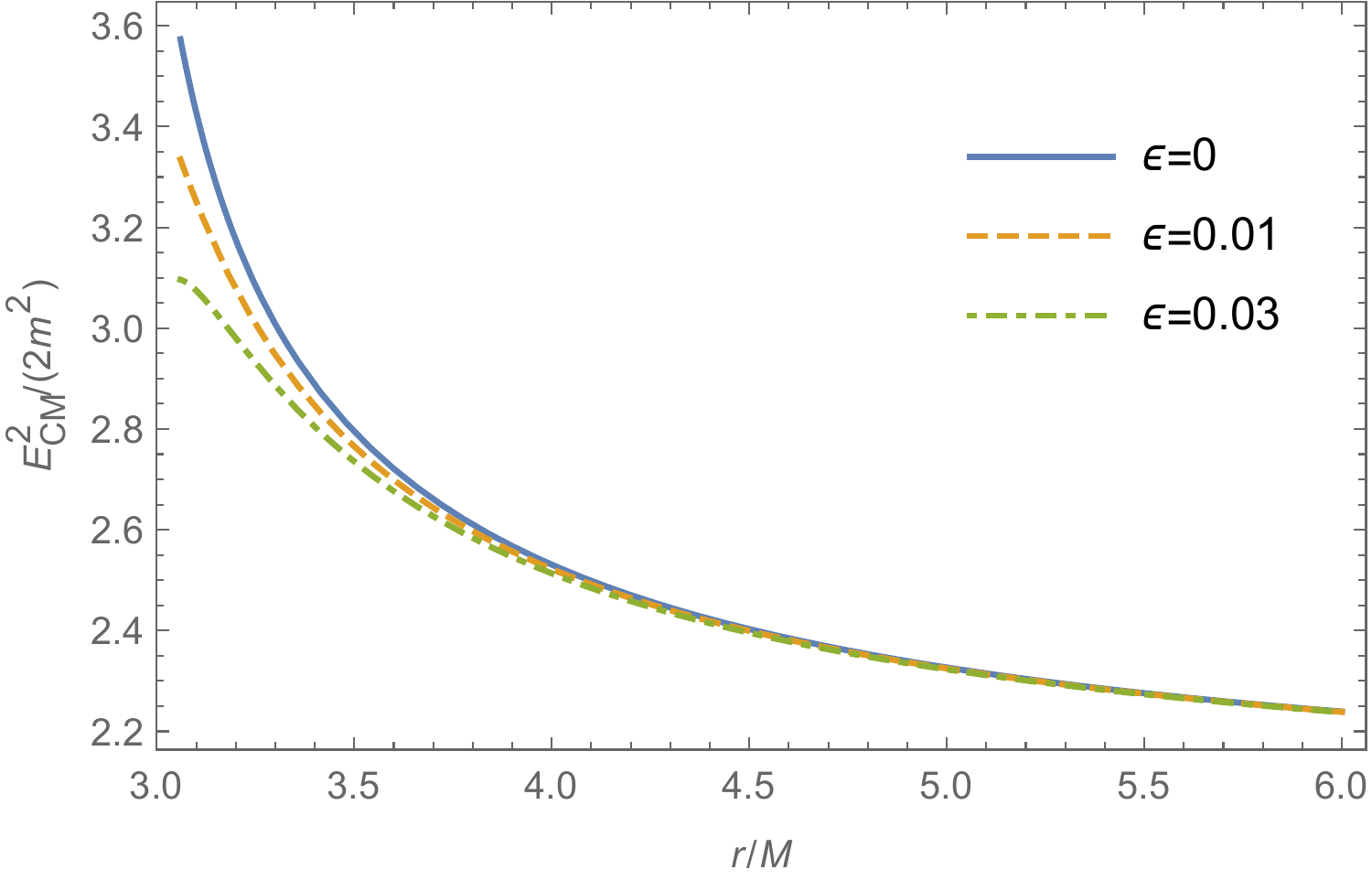}
\end{center}
\caption{The radial dependence of the CM energy of the colliding charged particle on the circular orbit with radially falling neutral one for the different values of the deviation parameter. The graph is plotted in the presence of the external uniform magnetic field with $b=0.25$.}\label{fig5}
\end{figure}

The radial dependence of the CM energy can be obtained using the expression (\ref{CME}). In Fig. \ref{fig5} it is shown how the CM energy of the two colliding particles behaves in the presence of an external uniform magnetic field and deviation parameter.

 It might be quite interesting to study the center of mass energy of the colliding particles in the case when the first particle, which assumed to be charged rotating at innermost stable circular orbit around compact quasi-Kerr object, is immersed in an external uniform magnetic field. Results obtained by using numerical calculus are presented in Tab.~\ref{3tab}.  It can be seen from the table that for the constant magnetic parameter the increase of deviation parameter $\epsilon$ increases the center of mass energy of the colliding particles as well which might seem a bit confusing since we saw that (see figures \ref{fig3} and \ref{fig5}) the increase of that parameter should reduce the center of mass energy. But it becomes clear when one remembers that the increase of $\epsilon$ reduces the ISCO radius and in the nearer regions the center of mass energy of particles rises exponentially outweighing the effect of the former one. { Similar to the case of extreme rotation of a compact object one can  deal with the magnitudes of the magnetic parameter $b$ which do not lead the ISCO radius to be very close to the regions of event horizon where the quasi-Kerr space-time metric does not behave well. From the Tab.~\ref{2tab} it is easy to see that the increase of the magnetic parameter reduces the ISCO radius which indicates that for the high values of parameter $b$ one could obtain ISCO radius in the region being  very close to the event horizon which restricts the usage of the linear approximation on small parameter $\epsilon$. However one can explore the case when the deviation parameter is absent $\epsilon=0$ as presented in the last three rows of the Tab.~\ref{3tab} which shows that for high values of magnetic parameter the center of mass energy becomes extremely large \cite{Frolov12}. }

\begin{table}
\caption{\label{3tab} Center of mass energy (more clearly $E^2_{cm}/2m^2$) of the two colliding particles at ISCO radius of the rotating charged particle . }
\begin{ruledtabular}
\begin{tabular}{ccccccccc}
 $\epsilon$ & -0.2 & -0.1 & 0 & 0.1 & 0.2 \\
\hline
$b=0$ & 4.7922 & 4.8093 & 4.8284 & 4.8498 & 4.8742
\\
\hline
$b=0.05$ & 4.7782 & 4.7946 & 4.8131 & 4.834 & 4.858
\\
\hline
$b=0.2$ & 5.0914 & 5.1196 & 5.1518 & 5.1893 & 5.2336
\\
\hline
$b=0.5$ & 5.5333 & 5.6001 & 5.678 & 5.7646 & 5.8261
\\
\hline

$b=10^3$ &  &  & 98.1725 &  & 
\\
\hline

$b=10^4$ &  &  & 305.953 &  & 
\\
\hline

$b=10^5$ &  &  & 963.123 &  & 
\\
\end{tabular}
\end{ruledtabular}
\end{table}

 From this section one can conclude that increase of the deviation parameter does not accelerate particles but slows them down, which has an opposite character to the magnetic field. The effects of the quasi-Kerr term of the metric become stronger in the near regions of the compact object while in the higher distances it behaves as a traditional Kerr black hole. This might play an important role for example when one attempts to identify if the object is a Kerr black hole or not.
\\

\section{Conclusion \label{sect5}}

In the present paper, we have investigated 
the explicit forms of the components of electromagnetic fields around a quasi-Kerr compact object
immersed in an asymptotically uniform magnetic field.
We have also investigated 
the motion of a charged test particle 
orbiting around a quasi-Kerr compact object 
immersed in an asymptotically uniform magnetic field. The main results obtained in this paper can be summarized as follows. 

We have obtained the exact analytic expressions for the electromagnetic field components around a quasi-Kerr compact object immersed in an external magnetic field. It was shown that at large distances the absolute values of the components of the electric field tend to zero while the components of the magnetic field tend to the corresponding values of $\sin\theta$ and $\cos\theta$ for chosen angles $\theta$. 

We analyzed the equation of motion of the charged particle motion around a quasi-Kerr compact object in a magnetic field. The analysis of the circular orbits of charged particles showed  that, increasing the value of the deformation parameter 
$\epsilon$ and the external magnetic field $B$ will reduce the ISCO radius. It is worth to note that as $\epsilon=0$, the results coincide with the results for a Kerr black hole as expected.


The measurements of the ISCO radius in accretion disks around compact objects can be used to obtain the constraint on the values of the deformation parameter $\epsilon$.
Observable properties of the accretion disc around black hole can be modeled using the spacetime metric and X-ray observation could give information about spacetime parameters, particularly if one compares with Kerr spacetime, one can get estimation of spin parameter~\cite{Steiner11,Gou14,McClintock14,Steiner10}. Comparison of the X-ray observations with the spacetime metric of alternative/modified theories gravity has been used to get constraint parameters of the spacetime metric~\cite{Cao18,Liu18}. In this paper we have obtained numerical results on ISCO around quasi-Kerr compact object which can be used to get rough constraint on $\epsilon$ parameter: using the error bar in the observation we can get rough constraint on the $\epsilon$ parameter. 
In~\cite{Cao18} the authors obtained an estimation for the spin parameter of selected X-ray sources with the confident of $90 \% $.  
 Taking into account that the observation of the ISCO radius will give maximum limit for errors of the measurements as 
 $10\%$ for the typical X-ray sources with the spin parameter estimated as $a/M=0.8 \pm 0.0008$ one can use this accuracy to get 
  rough constraints on $\epsilon$.
Since the ISCO radius decreases with an increase of the $\epsilon$ parameter, one may get rough estimation as $\epsilon \gtrsim -0.012$.
Note that a more detailed constraint on parameter $\epsilon$ can be obtained, e.g. using the continuum fitting method (see, e.g.\cite{Steiner11,Gou14,Cao18,Liu18} for the method review).
The results of the study the particles collision show that an increase of the deviation parameter does not accelerate particles but slows them down which has an opposite character to the magnetic field. The effects of the quasi-Kerr term of the metric become stronger in the near regions of the compact object while in the higher distances it behaves as a traditional Kerr black hole.

\begin{acknowledgments}

	This work was supported by the Innovation Program of the Shanghai Municipal Education Commission (Grant No.~2019-01-07-00-07-E00035), the National Natural Science Foundation of China (Grant No.~U1531117), and Fudan University (Grant No.~IDH1512060).
	B.N. also acknowledges support from the China Scholarship Council (CSC), grant No.~2018DFH009013.  
	The research is supported in part by Grant No. VA-FA-F-2-008 and No.YFA-Ftech-2018-8 of the Uzbekistan Ministry for Innovation Development, by the Abdus Salam International Centre for Theoretical Physics through Grant No. OEA-NT-01 and by Erasmus+ exchange grant between Silesian University in Opava and National University of Uzbekistan. A.A. thanks  Nazarbayev University for  hospitality. 
	
\end{acknowledgments}

\bibliographystyle{apsrev4-1}  
\bibliography{gravreferences}

\begin{appendix}
\section{ functions }\label{appendix}

\begin{widetext}

The functions introduced in (\ref{e1})--(\ref{b2}) have the following form:

\begin{eqnarray}
\mathcal{P}_1 &=& 2 r \Sigma  (\cos 2 \theta +3) \left(-2 a^2 M \sin ^2\theta +a^2 r+r^3\right)+8 a^2 M r^3 \sin ^2\theta  (\cos 2 \theta +3)
\\\nonumber
&& -\Sigma ^2
   \left(\left(a^2+3 r^2\right) \cos 2 \theta +3 a^2+r^2\right)
   \\\nonumber
   \\
\mathcal{P}_2 &=& a^2 M r (3 \cos 2 \theta +1)-2 \left(a^2+r^2\right) \Sigma
    \\\nonumber
    \\
\mathcal{P}_3 &=& 2 a^2 M r \sin ^2\theta -4 a^2 M r+\mathcal{R}
    \\\nonumber
    \\
\mathcal{P}_4 &=& a^2 M \Sigma  \sin ^2\theta -2 a^2 M r^2 \sin ^2\theta -4 a^2 M r^2 - 2 a^2 M \Sigma +r \Sigma ^2
\\\nonumber
\\
%
%
%
%
%
N_1 &=&\frac{M F_1 \left(3 \cos ^2\theta -1\right) (1-2 M/r)}{
\mathcal{R}} \left(-2 r (\cos 2 \theta +3) \Sigma \left(-2 a^2 M \sin ^2\theta +a^2
   r+r^3\right)\right.
   \\\nonumber
&&\left. -8 a^2 M r^3 \sin ^2\theta  (\cos 2 \theta +3)+\Sigma ^2 \left(\left(a^2+3 r^2\right) \cos 2 \theta +3 a^2+r^2\right)\right)
\\\nonumber
\\
N_2 &=& \frac{\Delta  F_1 M \Sigma ^2 (\cos 2 \theta +3) (3 \cos 2 \theta +1) \left(2 r^2-\Sigma \right)}{4 \mathcal{R} \left(1-2 M / r\right)}
\\\nonumber
\\
N_3 &=& \frac{F_1 M}{2}  \left(2 r^2 \Sigma  \left(-18 M^2-8 M r+7 r^2\right)+r \cos 2 \theta  \left(2 r \Sigma  \left(r^2-6 M^2\right)+8 M r^3 (2 M-r)+\Sigma
   ^2 (4 M-r)\right) \right.
\\\nonumber
 && \left.  +24 M r^4 (2 M-r)+7 r \Sigma ^2 (4 M-r)-4 \Sigma ^3\right)+\frac{(1-2 M/r)}{r} \left(F_2 M \sin ^2\theta  (2 M-r) \left(8 a^2 M r^3 \sin
   ^2\theta  \right. \right.
\\\nonumber
&& \left. \left.   -16 a^2 M r^3+2 r^2 \Sigma  \left(a^2+r^2\right)+\Sigma ^2 (a-r) (a+r)\right)+M r \Sigma  F_2' \sin ^2\theta  (2 M-r) \left(4 a^2 M
   r-\mathcal{R}\right) \right.
\\\nonumber
&& \left.  +r^3 F_1' \left(M r \Sigma  \left(\sin ^2\theta  (\Sigma -2 M r)+4 M r-4 \Sigma \right)+\Sigma (r)^3\right)\right)
\\\nonumber
\\
N_4 &=& \frac{\Delta  F_1 (\cos 2 \theta +3) (3 \cos 2 \theta +1) \left(a^2 M \left(2 r^2-\Sigma \right)+2 r \Sigma ^2\right)}{2 \Sigma  \mathcal{R}
   \left(1-\frac{2 M}{r}\right)}
\\\nonumber
\\
N_5 &=& \frac{(3 \cos 2 \theta +1)}{\Sigma ^3} \left(\frac{2 a^2 F_1 M r^4}{(r-2 M)^2} \left(r \Sigma  \left(-9 M^2-M r+2 r^2\right)+6 M r^3 (2 M-r) \right. \right.
\\\nonumber
&& \left. \left.  +M r \cos 2 \theta  \left(4 M
   r^2-3 M \Sigma -2 r^3+r \Sigma \right)+\Sigma ^2 (4 M-r)\right)-\frac{2 a^2 M r^5 \Sigma  F_1' }{4 M-2 r}(M r \cos 2 \theta +3 M r-2 \Sigma ) \right.
\\\nonumber
&& \left.  +r \Sigma  \mathcal{R} F_2' (r-2 M) \left(4 a^2 M r-\mathcal{R}\right)+2 F_2 (2 M-r) \left(2 a^4 M r \left(4 M r^3+\Sigma  \left(2 r^2+\Sigma
   \right)\right) \right. \right.
\\\nonumber
&& \left. \left. +a^2 M r \left(2 r^2-\Sigma \right) \left(2 r^2 \Sigma -\mathcal{R}\right)-a^2 M r \cos 2 \theta  \left(8 a^2 M r^3+\mathcal{R}
   \left(\Sigma -2 r^2\right)\right)-\Sigma  \mathcal{R} \left(\mathcal{R}-2 r^2 \Sigma \right)\right)\right)
   \\\nonumber
   \\
%
K_1 &=& -\frac{F_2 M \sin 2 \theta  (3 \cos 2 \theta +1) \left[a^2 M r (3 \cos 2 \theta +1)-2 \Sigma  \left(a^2+r^2\right)\right]}{\mathcal{R}}
\\\nonumber
\\
K_2 &=& F_2 M (2 M-r) \left[a^2 M r (28 \cos 2 \theta +9 \cos 4 \theta -5)-4 \Sigma  \left(a^2+r^2\right) (3 \cos 2 \theta -1)\right]
\\\nonumber
&&-\frac{F_1 \Delta  M
   r^3 \Sigma ^2 (3 \cos 2 \theta +1)}{\mathcal{R}}-4 F_1 r^2 (2 M r-\Sigma ) (M r (3 \cos 2 \theta +5)-3 \Sigma )
   \\\nonumber
   \\
K_3 &=& \frac{F_1 \sin 2 \theta  (3 \cos 2 \theta +1) \left(\Sigma  \left(a^2+r^2\right)-\Delta  M r\right) \left(\Sigma  \left(a^2+r^2\right)-4 a^2 M r \cos
   ^2\theta \right)}{2 \Sigma  \mathcal{R} \left(1-\frac{2 M}{r}\right)}
   \\\nonumber
   \\
K_4 &=& \frac{\frac{4 F_1 a^2 M^2 r^5 \sin ^2\theta  (9 \cos 2 \theta -1)}{2 M-r}+F_2 \mathcal{R} \left[-3 a^2 M r \cos 4 \theta +\cos 2 \theta \left(4
   a^2 M r+6 \mathcal{R}\right)+a^2 (-M) r-2 \mathcal{R}\right]}{2 \Sigma ^2}
   \\\nonumber
   \\
K_5 &=& \frac{a^2 M r \sin 2 \theta }{\Sigma ^2 (2 M-r)} \left[4 F_2 (r-2 M) \left[a^2 M r \sin ^2\theta  (9 \cos 2 \theta -1)+\Sigma  \left(a^2+r^2\right) (3 \cos 2 \theta
   -1)\right] \right.
\\\nonumber
&&  \left. -4 F_1 r^3 (3 \cos 2 \theta -1) (2 M r-\Sigma )\right]
\end{eqnarray}
\begin{eqnarray}
R_1 &=& \frac{B F_1 \sin 2 \theta (3 \cos 2 \theta +1) (1-2 M/r) \sqrt{\frac{\Sigma }{\Delta }} \left(a^2 M r \cos 2 \theta +3 a^2 M
   r-\mathcal{R}\right)}{8 \sqrt{\mathcal{Q}}}
   \\\nonumber
   \\
R_2 &=& \frac{\Delta ^2 F_1 \Sigma  \left(3 \cos ^2\theta -1\right)}{2 \sqrt{\mathcal{Q}} \mathcal{R} \left(1-\frac{2 M}{r}\right)}
\\\nonumber
\\
R_3 &=& \left(1- 3 \cos ^2\theta \right) \left[F_1 r \left[2 \Delta  M r^3+\Sigma ^2 (r-2 M)^2-\Delta  r^2 \Sigma \right] \right.
\\\nonumber
&& \left. -F_2 \Delta  (r-2 M) \left[-a^2 M r
   \cos 2 \theta +a^2 (M r+\Sigma )+\Sigma  \left(r^2+\Sigma \right)\right]\right]
   \\\nonumber
   \\
R_4 &=& \frac{2 F_1 a^2 M^2 r^4 \Sigma ^2 \sin ^2\theta  (1-9 \cos 2 \theta )}{\Sigma ^4 (1- 2 M)}- \frac{F_2 \mathcal{R}}{\Sigma ^4} \left[\Sigma ^2 (\mathcal{R}-3 \mathcal{R} \cos 2
   \theta )-2 a^2 M r \Sigma ^2 \sin ^2\theta  (3 \cos 2 \theta +1)\right]
   \\\nonumber
   \\
R_5 &=& -\frac{4 F_2 a^2 M \sin 2 \theta  }{r \Sigma ^2 } \left(a^2 M r \sin ^2\theta  (9 \cos 2 \theta -1)+\Sigma  \left(a^2+r^2\right) (3 \cos 2 \theta
   -1)\right)
\\\nonumber
&&   -\frac{4 a^2 M \sin 2 \theta F_1 r^3 }{r \Sigma ^2 (2 M-r)}  (3 \cos 2 \theta -1) (2 M r-\Sigma )
\end{eqnarray}
\begin{eqnarray}
D_1 &=& \frac{B F_2 \Delta  \sin^2 \theta  \left(1-3 \cos ^2\theta \right) \left[-a^2 M \Sigma  (\cos 2 \theta +3)+2 a^2 M r^2 (\cos 2 \theta +3)+2 r
   \Sigma ^2\right]}{4 r^2 \sqrt{\mathcal{Q} \Sigma } }
   \\\nonumber
   \\
D_2 &=& \frac{F_1 \Delta ^2  r \Sigma  \left(1-3 \cos ^2\theta \right)}{2 \mathcal{R} (2 M-r) \sqrt{\mathcal{Q}}}
   \\\nonumber
   \\
D_3 &=& \frac{  F_2 \Delta \left(1-3 \cos ^2\theta \right) }{r^2  \sqrt{\mathcal{Q} \Sigma }} \left[2 a^2 M r \sin ^2 \theta +\Sigma  \left(a^2+r^2+\Sigma \right)\right]
\\\nonumber
&&+ \frac{F_1 r}{r^3 (1- 2 M/r) \sqrt{\mathcal{Q} \Sigma }}
   \left[\Delta  r^2 (\Sigma -2 M r)-\Sigma ^2 (r-2 M)^2\right]
   \\\nonumber
   \\
D_4 &=& \frac{\Delta  (3 \cos 2 \theta +1)}{\Sigma ^3}
\\\nonumber
&&\times \Bigl[\frac{r \Sigma  \left(4 a^2 M^2 r^5 \sin ^2\theta  F_1'+\mathcal{R}^2 (2 M-r) F_2'\right)}{4 M-2 r}-\frac{4 a^2 F_1 M^2 r^5 \sin ^2\theta  \left(2 r^2 (2 M-r)+\Sigma  (r-3 M)\right)}{(r-2 M)^2} \Bigr.
\\\nonumber
&& \Bigr. -F_2 \mathcal{R} \left[4 a^2 M r^3 \sin ^2\theta +\Sigma  \left(a^2 M r \cos 2 \theta +a^2 (-M) r-2 r^2 \Sigma +\mathcal{R}\right)\right]\Bigr]
\\\nonumber
\\
D_5 &=& \frac{2 a^2 M r (3 \cos 2 \theta +1)}{\Sigma ^3 (r-2 M)^2} \left[F_1 r^3 \left(\Sigma  \left(12 M^2 r+\Sigma  (r-4 M)-2 r^3\right)+8 M r^3 (r-2 M)\right) \right.
\\\nonumber
&& \left. -(2 M-r) \left(F_2 (2
   M-r) \left(8 a^2 M r^3 \sin ^2\theta +2 r^2 \Sigma  \left(a^2+r^2\right)+\Sigma ^2 (a-r) (a+r)\right) \right. \right.
\\\nonumber
&& \left. \left. +r \Sigma  \left(r^3 F_1' (\Sigma -2 M
   r)+\mathcal{R} F_2' (r-2 M)\right)\right)\right]
\end{eqnarray}
\end{widetext}
where $F_1'$ with $F_2'$ are derivatives of the functions over $r$.
\end{appendix}
\end{document}